\documentclass[iop]{emulateapj}

\usepackage{amsmath, amsthm, amssymb, amsfonts}
\usepackage{hyperref}
\usepackage{upgreek}
\usepackage{booktabs}
\hypersetup{
	colorlinks=true,
	linkcolor=black,
	filecolor=black,
	urlcolor=black,
	citecolor = blue,
}
\usepackage{graphicx, nicefrac}

\slugcomment{FINAL VERSION October 17, 2023}

\shorttitle{Why do SAMs predict higher scatter in the SMHMR than hydrodynamical simulations?}
\shortauthors{Porras-Valverde et al.}

\begin{document}



\title{Why do semi-analytic models predict higher scatter in the stellar mass--halo mass relation than cosmological hydrodynamic simulations?}


\author{Antonio J. Porras-Valverde\altaffilmark{1,2}, John C. Forbes\altaffilmark{3,4}, Rachel S. Somerville\altaffilmark{3}, Adam R. H. Stevens\altaffilmark{5}, Kelly Holley-Bockelmann\altaffilmark{1,6}, Andreas A. Berlind\altaffilmark{1}, and Shy Genel\altaffilmark{3}}
\affil{\altaffilmark{1}Department of Physics and Astronomy, Vanderbilt University, 6301 Stevenson Science Center, Nashville, TN 37212, USA \\
\altaffilmark{2}Department of Astronomy, Yale University, P.O. Box 208101, New Haven, CT 06520, USA \\
\altaffilmark{3}Center for Computational Astrophysics, Flatiron Institute, 162 5th Ave, New York, NY 10010, USA\\
\altaffilmark{4}School of Physical and Chemical Sciences--Te Kura Mat\=u, University of Canterbury, Christchurch 8140, New Zealand\\
\altaffilmark{5}International Centre for Radio Astronomy Research, The University of Western Australia, 7 Fairway, Crawley, WA 6009, Australia \\
\altaffilmark{6}Department of Life and Physical Sciences, Fisk University, 1000 17th Avenue N., Nashville, TN, 37208, USA}







\begin{abstract}

Semi-analytic models (SAMs) systematically predict higher stellar-mass scatter at a given halo mass than hydrodynamical simulations and most empirical models. Our goal is to investigate the physical origin of this scatter by exploring modifications to the physics in the SAM {\sc Dark Sage}. We design two black hole formation models that approximate results from the {\sc IllustrisTNG 300-1} hydrodynamical simulation. In the first model, we assign a fixed black hole mass of $10^{6}\, \mathrm{M}_{\odot}$ to every halo that reaches $10^{10.5}\, \mathrm{M}_{\odot}$. In the second model, we disregard any black hole growth as implemented in the standard {\sc Dark Sage} model. Instead, we force all black hole masses to follow the median black hole mass - halo mass relation in {\sc IllustrisTNG 300-1} with a fixed scatter. We find that each  model on its own does not significantly reduce the scatter in stellar mass. To do this, we replace the native Dark Sage AGN
feedback model with a simple model where we turn
off cooling for galaxies with black hole masses above $10^{8}\, \mathrm{M}_{\odot}$. With this additional modification, the SMBH seeding and fixed conditional distribution models find a significant reduction in the scatter in stellar mass at halo masses between $10^{11-14}\, \mathrm{M}_{\odot}$. These results suggest that AGN feedback in SAMs acts in a qualitatively different way than feedback implemented in cosmological simulations. Either or both may require substantial modification to match the empirically inferred scatter in the Stellar Mass Halo Mass Relation (SMHMR).

\end{abstract}





\section{Introduction}\label{introduction}

The standard model of cosmology, $\Lambda$CDM, is the backbone for our understanding of galaxy formation \citep{1984Natur.311..517B, Cen1994X-rayUniverse, Cen1999COSMICEVOLUTION, Cen1999ACCURACYCORRECTIONS}. Within this model, collapsed dark matter halos forming along filaments in the large-scale structure of the Universe serve as the skeleton to the luminous baryonic matter that accumulates in galaxies \citep{White1978, 1993MNRAS.264..201K}. Dark matter halos are the basis of essentially all simplified physical modeling of galaxy formation, but their properties are difficult to measure directly.

For decades, \textit{empirical} and \textit{physical} models have been used to study the multivariate distributions of properties formed and evolved in galaxies and halos. \textit{Empirical models} constrain the relationship between halos and galaxies by assigning galaxies to halos using a variety of techniques that differ in complexity such as abundance matching \citep{1999ApJ...523...32C}, halo occupation distributions (HOD) \citep{2004ApJ...609...35K}, conditional luminosity functions \citep{2003MNRAS.339.1057Y}, and models that flexibly connect galaxies over time to dark matter halo growth histories \citep{2009ApJ...696..620C, 2010ApJ...710..903M, Hearin2013TheColour, Behroozi2019Universemachine:010}. \textit{Physical models} (see \citet{2015ARA&A..53...51S} for a review) directly evolve galaxies, or more generally gas with an explicit parameterization of the physics of galaxy formation. Physical models can be further divided into semi-empirical or semi-analytical models \citep[e.g.][]{Croton2006, Croton2016, Somerville2001TheGalaxies, Benson2012, Henriques2015, Stevens2016}, and hydrodynamical simulations \citep[e.g.][]{2015MNRAS.450.1349K, 2016A&C....15...72M, Naiman2018FirstEuropium}. SAMs are generally built on physically and/or empirically motivated relations of galaxy properties that are described macroscopically, whereas hydrodynamical simulations self-consistently evolve galaxies in tandem with the large-scale structure of the simulated Universe, albeit with parameterized models for processes below the resolution limit. One key prediction from \textit{empirical models}, the mean stellar mass-to-halo mass relation (SMHMR) is frequently used explicitly to calibrate the parameters of \textit{physical models} \citep[e.g.][]{Stevens2016, 2018MNRAS.473.4077P, 2019MNRAS.487.3581F}. While the mean relation is calibrated, the scatter is not, so the scatter at fixed halo mass produced in physical models is a genuine prediction of most physical models.

The review on the galaxy--halo connection by \citet{Wechsler2018} highlights the consensus ratio of stellar mass to halo mass as a function of halo mass among empirical models for central galaxies at $z = 0$. Quantitative inconsistencies can still arise between empirical models from different stellar mass functions, luminosity functions, and/or halo mass functions, but, qualitatively, essentially every model agrees on the main features. Among these features is the peak around $M_h = 10^{12}\, \mathrm{M}_{\odot}$, which corresponds to a few percent of the matter in galaxies being composed of stars. This means that if these halos have a baryon fraction of about 17 percent, only about 20--30 percent of the baryons have turned into stars. For galaxies above or below this peak, the stars formed per unit dark matter mass (sometimes called the galaxy formation efficiency) drops. This drop in efficiency comes from feedback mechanisms that affect star formation and black hole growth, which affect galaxies differently based on their mass. For massive galaxies ($M_h \sim 10^{12-15}\, \mathrm{M}_{\odot}$), feedback from supermassive black holes suppresses star formation through ejection of cold gas from the disk and heating up the hot gas surrounding the galaxy \citep{2006MNRAS.370..645B, Croton2006, 2008MNRAS.391..481S}. For low-mass galaxies ($M_h \sim 10^{9-11}\, \mathrm{M}_{\odot}$), stellar feedback is more efficient at preventing the formation of stars through some combination of driving cold gas out of the galaxy \citep[e.g.][]{dekel1986, 2012MNRAS.421.3522H}, slowing the formation of stars \citep{forbes2016}, or preventing the accretion of new gas \citep[e.g.][]{Lu2015, Pandya2022, Carr2023}. For even smaller size galaxies ($M_h < 10^{9}\, \mathrm{M}_{\odot}$), the gravitational potential is too shallow to combat photoionization feedback \citep{2000ApJ...539..517B}. 

Most \textit{empirical models} fundamentally assume that the scatter in stellar mass at fixed halo mass ($\sigma(\log_{10}(M_* | M_h))$) is irreducible. In abundance matching models, the scatter is introduced as a parameter in the model, representing the width of the distribution of galaxies around the mean galaxy-halo relation, assumed to have a lognormal distribution \citep{2010ApJ...717..379B}. It is constrained by comparing the model predictions to observed properties of the galaxy distribution, e.g. the stellar mass function, and clustering. The difficulty of constraining the stellar mass scatter using abundance matching models is shown in \citet{Wechsler2018}, who show that models that match the SDSS stellar mass function yield different stellar mass scatter values. In HOD models that use luminosity or stellar mass thresholds, the scatter is added in the form of the central galaxy occupation function. This function determines the probability of having a galaxy above the threshold based on the halo mass, considering that there can only be one or no central galaxies. The scatter inferred from HOD models is linked to the scatter in halo mass at a fixed galaxy mass (or luminosity). Because the intrinsic scatter is degenerate, it is difficult to constrain if, for example, halo properties beyond mass are not accounted for. Since HODs commonly assume galaxy occupation based on halo mass, the influence of secondary properties like halo concentration or age is neglected. This ignores the potential impact of assembly bias, as galaxy clustering might be affected by other factors such as the halo formation history, star formation histories, or galaxy intrinsic properties \citep{2005MNRAS.363L..66G, 2006ApJ...652...71W, 2007MNRAS.374.1303C, 2018MNRAS.475.4411S, 2018ApJ...853...84Z, 2019MNRAS.486..582P, 2019MNRAS.484.1133C, 2020MNRAS.493.5506H, 2021MNRAS.501.1603H}.

In \textit{physical models} however, one can distinguish between irreducible scatter and the scatter owing to other galaxy properties. 
In physics-based models, scatter in stellar mass likely arises from one or more physical processes such as halo assembly history, feedback, and environmental effects. Halos of the same mass can form through different assembly histories, which causes a variation in the gas content and stars formed in each galaxy. \citet{2022MNRAS.517.6091G} find a correlation between the residuals from the median SMHMR for individual halos in {\sc TNG} and the {\sc Santa Cruz SAM}, suggesting that the scatter in the SMHMR is set by similar processes in the two models. \citet{2017MNRAS.465.2381M} determine that for $M_h < 10^{12.5}\, \mathrm{M}_{\odot}$, $0.03\ \mathrm{dex}$ of the scatter in stellar mass can be attributed to halo concentration, but for $M_h > 10^{12.5}\, \mathrm{M}_{\odot}$, incorporating concentration does not lead to a reduction in the stellar mass scatter, suggesting that other distinct physical processes are influencing the scatter.

\citet{2017MNRAS.467.3533T} employ a model to investigate the effects of various quenching mechanisms on the scatter in stellar mass. They find that stellar mass quenching is the only model that consistently predicts a $0.2\ \mathrm{dex}$ scatter assuming the quenching threshold is constant with time. To reduce the scatter in their models which predicted a larger scatter, significant correlations between galaxy properties, such as metallicity or mean stellar age, and halo formation history would be necessary. \citet{2019MNRAS.484..915M} find that the scatter in star formation rate at fixed stellar mass arises from a mix of short time-scale fluctuations (spanning 0.2–2 Gyr), likely linked to self-regulation from cooling, star formation, and outflows. However, the primary contributor to the scatter is the variations on long time-scales (about 10 Gyr) associated with differences in halo formation times.

Observationally, the SMHMR in the local Universe is constrained using different methods such as: non-parametric abundance matching \citep{2010ApJ...717..379B, 2013ApJ...771...30R, 2013ApJ...770...57B}, parameterized abundance matching, in which the SMHMR is parameterized, so that those parameters are fit with the observed stellar mass function and other observed relations \citep{2010MNRAS.404.1111G, 2010MNRAS.402.1796W, 2010ApJ...710..903M, 2013MNRAS.428.3121M, 2018MNRAS.477.1822M}, from modeling HODs \citep{2007ApJ...667..760Z}, the conditional luminosity function \citep{2012ApJ...752...41Y}, and by direct measurement of X-ray clusters \citep{2004ApJ...617..879L, 2009ApJ...695..900Y, 2009ApJ...699.1333H, Kravtsov2018StellarHalos}. The observations used to refine these methods are the stellar mass function at a specific time period, stellar mass, star formation rates over time, group stellar mass function, the conditional galaxy stellar mass function within groups, galaxy clustering, galaxy–galaxy lensing, and satellite kinematics. The main disagreement among these techniques in regards to the mean SMHMR comes from systematic discrepancies in measuring the stellar mass at the high mass end. For low mass galaxies, the uncertainties are driven by small, or incomplete sample statistics.

Using halo bias, \citet{2017ApJ...839..121T} determined that a scatter model with value of $\sigma(\log_{10}(M_* | M_h)) = 0.18\ \mathrm{dex}$ from their SMHMR prediction of the Baryon Oscillation Spectroscopic Survey (BOSS) from the Sloan Digital Sky Survey (SDSS) galaxies at $z = 0.5$ fits best for galaxies with stellar masses $M_* > 10^{11}\,{\rm M}_\odot$. Observations of galaxy clustering \citep{2012ApJ...744..159L, 2015MNRAS.449.1352C}, galaxy--galaxy lensing \citep{2006MNRAS.368..715M, 2015MNRAS.447..298H}, and satellite kinematics \citep{2007ApJ...654..153C, 2011MNRAS.410..210M} have been used to constrain the scatter in stellar mass. \citet{Zu2015MappingDR7} found scatter values of $\sigma(\log_{10}(M_* | M_h)) = 0.18\ \mathrm{dex}$ at halo masses of $10^{14}\, \mathrm{M}_{\odot}$ and $\sigma(\log_{10}(M_* | M_h)) = 0.22\ \mathrm{dex}$ at halo masses of $10^{12}\, \mathrm{M}_{\odot}$. Estimating halo masses from X-ray measurements, \citet{Kravtsov2018StellarHalos} find $\sigma(\log_{10}(M_* | M_h)) = 0.17\ \mathrm{dex}$ from a direct measurement of the scatter in the mean SMHMR at halo masses between $10^{14}-10^{15}\, \mathrm{M}_{\odot}$. Additionally, tracing the gravitational potential of central galaxies using satellite kinematics, \citet{2011MNRAS.410..210M} finds $\sigma(\log_{10}(M_* | M_h)) < 0.2\ \mathrm{dex}$ for galaxies with stellar masses between $10^{10}-10^{12}\, \mathrm{M}_{\odot}$, results that are in agreement with galaxy clustering and galaxy lensing results. All in all, the scatter is generally constrained to be $\la 0.2\ \mathrm{dex}$ for central galaxies ($z = 0$) with $M_h > 10^{11}\, \mathrm{M}_{\odot}$. Understanding the causes and implications of the scatter in the SMHMR is crucial for developing a more complete understanding of galaxy formation.

In this paper, we investigate the physical origin of the scatter in the stellar mass--halo mass relation for massive halos by exploring modifications to the physics in the SAM {\sc Dark Sage}. We focus our studies on $M_h > 10^{11.0}\, \mathrm{M}_{\odot}$ because the scatter there is constrained by a variety of observations \citep{2011MNRAS.410..210M, Zu2015MappingDR7, Kravtsov2018StellarHalos}. We find compelling evidence that much of the scatter arises from the treatment of AGN quenching, so our modifications focus on this area of the SAM. Within {\sc Dark Sage}, we isolate two potential contributions to the stellar mass scatter at fixed halo mass: 1) the black hole mass distribution and 2) the treatment of AGN feedback. To explore the former, we force {\sc Dark Sage} to follow the black hole mass--halo mass median relation produced by the {\sc IllustrisTNG 300-1} hydrodynamical simulation using the highest-resolution box ({\sc TNG} hereafter). To explore the latter, we run {\sc Dark Sage} with AGN quenching models that affect the various gas reservoirs in the SAM differently. 

This paper is organized as follows. In Section \ref{Methodology_DS}, we present an overview of the semi-analytic model {\sc Dark Sage}, where we detail how black hole seeding and AGN feedback is implemented in the model. We do the same in Section \ref{Methodology_TNG} for the {\sc TNG} simulation. In Section \ref{SMHMR_section}, we examine the stellar mass-to-halo mass relation in both models. Section \ref{BHM_seedingAGN} explores the black hole seeding and AGN effect in changing the $\sigma(\log_{10}(M_* | M_h))$. Sections \ref{discussion} and \ref{conclusion} discuss our results and present conclusions. The simulation uses cosmological parameters from the Wilkinson Microwave Anisotropy Probe data \citep{Spergel2003}, where $\Omega_M = 0.25$, $\Omega_{\Lambda} = 0.75$, $\Omega_b = 0.045$, $\sigma_8 = 0.9$, and $h = 0.73$. These parameters are not the same used by {\sc TNG}. For information about {\sc TNG} cosmological parameters, see section \ref{Methodology_TNG}.

  

\section{Semi-analytic Model: {\sc Dark Sage}}\label{Methodology_DS}

The semi-analytic model {\sc Dark Sage}'s \citep{Stevens2016} framework comes from {\sc SAGE} \citep{Croton2016}. The focus on tracing the evolution of the galactic disk through annular structures is what makes {\sc Dark Sage} stand out from other semi-analytic models. {\sc Dark Sage} self-consistently evolves the one dimensional structure of disks in galaxies using 30 equally-spaced logarithmic bins of fixed angular momentum, a method inspired by \citet{Fu2010, Fu2013}. Most SAMs trace the physics within the disk as global quantities, but {\sc Dark Sage} is able to calculate local processes within annuli such as star formation rates, disk instabilities, AGN energy dissipation, and so forth \citep[see also][]{Forbes2012,Forbes2014a,2019MNRAS.487.3581F,Henriques2020}. In {\sc Dark Sage}, galaxies are formed within halos of hot gas that cool down through radiative cooling and condensation of hot gas \citep{White1978}. Once it cools down, the gas collapses to form a rotationally-supported disk. {\sc Dark Sage} uses two annular disk structures, for stars and gas, to calculate physical processes. The stellar and gas disks may also be misaligned due to precession, which is a natural consequence of having a rotating fluid body in an axisymmetric potential where their axes of symmetry are offset. In fact, the gas disk precesses to becomes parallel to the stellar spin axis until it reaches coplanarity. {\sc Dark Sage} galaxies resolve \citet{Toomre1964} instabilities by transferring unstable gas or stars to neighboring annuli such that angular momentum is conserved. In some cases, an unstable annulus may experience a starburst to resolve instabilities. However, if instabilities reach the innermost annulus, unstable stars are added to the instability-driven bulge, whereas unstable gas is added directly into the central black hole. In addition, {\sc Dark Sage} allows stellar disks to be completely destroyed if galaxies have a major merger. The resulting merged galaxy will subsequently acquire a newly-formed stellar disk structure. In the case of a minor merger, the stellar disk of the satellite is directly added into the merger-driven bulge. 

In both major and minor mergers, the gas from the secondary goes to the primary's gas disk annuli of the respective specific angular momentum. Before determining how much gas is added into the primary's gas disk, a fraction ($f_{\rm BH}$) of the gas feeds directly the central black hole of the primary (see equation \ref{eq:BH_quasar}). Once that happens, after the gas of the secondary goes to the respective specific angular momentum annulus, the annulus experiences a merger-driven starburst. Both major and minor mergers are important processes that may trigger AGN and stellar feedback activity. The careful treatments within the disk structures of galaxies are what makes {\sc Dark Sage} an excellent tool to study star formation quenching and galactic angular momentum.

In this paper, we use the 2018 version of {\sc Dark Sage} \citep{Stevens2018ConnectingSAGE}. 
In this version, {\sc Dark Sage} updates the way in which the cooling scale radius for the \textit{hot-mode} cooling is done by the fitting function from \citet{Stevens2017HowEAGLE}. 
The model uses merger trees from the Millennium suite, a dark matter-only simulation with a boxsize of 500 $h^{-1}$ Mpc  and particle mass resolution of 8.6$\times 10^8 \ h^{-1}\, \mathrm{M}_{\odot}$~\citep{Springel2005Nat}.  The simulation uses cosmological parameters from the Wilkinson Microwave Anisotropy Probe data \citep{Spergel2003}, where $\Omega_M = 0.25$, $\Omega_{\Lambda} = 0.75$, $\Omega_b = 0.045$, $\sigma_8 = 0.9$, and $h = 0.73$.  To ensure that galaxies at redshift 0 live in well-resolved halos, we adopt a halo mass cut of $10^{11.2}\, \mathrm{M}_{\odot}$; our {\sc Dark Sage} sample contains galaxies within the stellar mass range of $10^{9}-10^{12}\, \mathrm{M}_{\odot}$. For a more detailed description of {\sc Dark Sage}, please refer to \citet{Stevens2016}.

\subsection{Black hole growth in {\sc Dark Sage}}\label{seedingBH_DS}

{\sc Dark Sage} seeds a zero-mass black hole at every newly-formed halo. Thereafter, black holes grow via four main channels: gas accretion from \textit{in-situ} unstable gas, gas accretion from galaxy mergers, mergers with other black holes, and accretion from hot halo gas. Now, let's examine each of these channels in some detail. As galaxies form and evolve, black holes begin acquiring mass by \textit{cold mode} gas accretion through Toomre disk instablities and galaxy mergers, and later by \textit{hot mode} gas accretion directly from the circumgalactic/intrahalo medium. Gas accretion from the hot halo is modeled using a Bondi--Hoyle--Lyttleton model \citep{Bondi1952} (see equation \ref{eq:BH_Bondi} below). Additionally, during major mergers, both black holes are simply added up. Gas may also funnel into the galactic center on short time-scales, which results in high-accretion rates and rapid black hole growth. The cold gas disk of the primary and secondary galaxy is summed up. A fraction of this summed up gas is directly accreted onto the black hole following equation \ref{eq:BH_quasar}, adapted from \citet{2000MNRAS.311..576K}, applied individually to each annulus:
\begin{equation}
\label{eq:BH_quasar}
\begin{aligned}
\Delta m_{\mathrm{cold}} = f_{\mathrm{BH}} \left[ 1 + \left( \frac{280 \ \mathrm{km s^{-1}}}{V_{\mathrm{vir}}} \right)^{2} \right]^{-1} \\ \sum_{i=1}^{30} \left( m_{\mathrm{i,cen}} + m_{\mathrm{i,sat}} \right) \min \left( \frac{m_{\mathrm{i,sat}}}{m_{\mathrm{i,cen}}}, \frac{m_{\mathrm{i,cen}}}{m_{\mathrm{i,sat}}}  \right),
\end{aligned}
\end{equation}
where the cold gas mass accreted onto the black hole via the disk is expressed by $\Delta m_{\mathrm{cold}}$. $f_{\mathrm{BH}}$ is the accretion efficiency that controls the fraction of cold gas accretion coming from major/minor mergers. $m_{\mathrm{i}}$ is the gas mass in annulus $i$ of the central or satellite. Note that most of the gas accreted by black holes has low specific angular momentum, similar to what is found in \citet{2018ApJ...860...20S}. In {\sc Dark Sage}, equation \ref{eq:BH_quasar} contributes the most to the black hole mass buildup. 

After gas is funneled directly into the galactic center, the rest of the cold gas in each annulus is subject to a merger-driven starburst phase following \citet{Somerville2001TheGalaxies}. Any leftover cold gas that is gravitationally unstable is then subject to {\sc Dark Sage}'s generic prescription for handling unstable gas as briefly described in the previous subsection. The movement of unstable gas may trigger instabilities elsewhere in the disk, sometimes cascading until unstable gas reaches the innermost annulus, where a fraction is then transferred directly into the black hole. 

\subsection{AGN feedback in {\sc Dark Sage}}\label{AGNfeedback_DS}

{\sc Dark Sage} uses \textit{quasar mode} and \textit{radio mode} feedback, each of which is tied to a different type of black hole growth. The \textit{quasar mode} feedback couples cold gas accretion to the interstellar medium (ISM), while hot gas accretion is coupled to the circumgalactic medium (CGM) via \textit{radio mode} feedback. Both feedback prescriptions scale with black hole mass (i.e. the more massive the black hole, the stronger the feedback). 

\textit{Radio mode} feedback dumps energy into the halo to prevent hot gas from cooling. This feedback mode is implemented at every timestep in the code. 
Without \textit{radio mode} feedback, overall, galaxies would accumulate large cold gas reservoirs and the subsequent high star formation rates would result in galaxies with higher stellar masses than observed in the real Universe (as demonstrated by \citealt{Croton2006}; also see the violet line in our Figure \ref{fig:SMHMR_plot}). {\sc Dark Sage} uses an identical \textit{radio mode} prescription as {\sc SAGE} \citep{Croton2016}, a model that implemented a \textit{radio mode efficiency} parameter, $\kappa_{\mathrm{R}}$, to control for the strength of the black hole accretion rate. The Bondi accretion rate is written in equation \ref{eq:BH_Bondi} as:
\begin{equation}
\dot{m}_{\mathrm{Bondi,R}}=\kappa_R \ \frac{15}{16} \uppi \mathrm{G} \ \upmu m_p \ \frac{kT}{\Lambda} m_{\mathrm{BH}},\label{eq:BH_Bondi}
\end{equation}
where $G$ is the gravitational constant, $\upmu$ is the mean molecular weight of the gas, $m_p$ is the mass of the proton, $k$ is the Stefan-Boltzmann constant, and $\Lambda$ is the cooling function of the gas, which describes how efficiently the gas can radiate energy away. 
$T$, is the temperature of the hot gas, which is assumed to be the halo's virial temperature, and $m_{\mathrm{BH}}$ is the black hole mass. Using this accretion rate, the luminosity of the black hole is estimated to be $L_{\mathrm{BH,R}} = \eta\, \dot{m}_{\mathrm{Bondi,R}}\, c^2$, where $\eta=0.1$ is the nominal radiative mass efficiency. This is used to estimate the amount of hot gas per unit time prevented from cooling:
\begin{equation}
\dot{m}_{\mathrm{heat}}= \frac{2\,L_{\mathrm{BH,R}}}{ V_{\mathrm{vir}}^2}.\label{eq:heating_rate}
\end{equation}
In {\sc Dark Sage}, \textit{radio mode} feedback is implemented via a heating radius $r_{\rm heat}$, interior to which gas cannot cool. First, to find $r_{\rm heat}$, {\sc Dark Sage} calculates the radius at which the energy gained through \textit{radio mode} feedback is balanced by the energy lost through gas cooling. This radius is only allowed to move outwards to retain information of past heating episodes. Second, to obtain the cooling radius $r_{\rm cool}$, {\sc Dark Sage} finds the radius at which the gas cooling time is equal to the halo dynamical time. Lastly, to calculate the new cooling rate, {\sc Dark Sage} uses the hot gas between $r_{\rm heat}$ and $r_{\rm cool}$.

The change in black hole mass due to \textit{quasar mode} accretion is given by equation \ref{eq:BH_quasar}. {\sc Dark Sage} calculates the energy budget from \textit{quasar mode} feedback as:
\begin{equation}
E_{\mathrm{BH,Q}} = \kappa_{\mathrm{Q}} \ \frac{1}{2} \ \eta \ \Delta m_{\mathrm{BH,Q}} \ c^2 ,
\label{eq:quasar_mode_totenergy}
\end{equation}
where $\kappa_{\mathrm{Q}}$ is the quasar mode efficiency. This energy is compared to the energy necessary to heat each successive annulus of cold gas, $E_{\mathrm{cold},i} = (1/2) m_i V_\mathrm{vir}^2$. If $E_\mathrm{BH,Q}$ (minus the energy required to heat interior annuli) exceeds $E_{\mathrm{cold},i}$, that annulus is heated. If energy remains after the entire cold gas disk is heated by the quasar mode, hot gas can also be ejected from the halo.

\section{Hydrodynamical simulation: {\sc TNG}}\label{Methodology_TNG}

The {\sc IllustrisTNG} is a state-of-the-art hydrodynamical simulation that models the evolution of galaxies and their environments in a cosmological context \citep{Pillepich2018FirstGalaxies, Springel2018FirstClustering, 2019MNRAS.490.3234N, Naiman2018FirstEuropium, Nelson2018FirstBimodality, Marinacci2018FirstFields}. The simulation was developed as an upgrade to the original Illustris simulation \citep{2014Natur.509..177V, 2014MNRAS.444.1518V, 2014MNRAS.445..175G, 2015MNRAS.452..575S, Nelson2015}. {\sc TNG} has a set of physical processes such as gas cooling and heating, star formation, stellar feedback, black hole growth, and feedback from active galactic nuclei. The simulation has been used to study a wide range of topics, including the role of feedback in shaping the properties of galaxies \citep{10.1093/mnras/sty1733, 2020MNRAS.493.1888T}, the connection between the properties of galaxies and their dark matter halos \citep{2019MNRAS.490.5693B, 2020MNRAS.491.5747M, 2020MNRAS.492.1671Z}, and the properties of galaxy clusters and their galaxies \citep{2019ApJ...876...82N, 2020MNRAS.494.1848S, 2020MNRAS.496.2673J}. {\sc TNG300-1} uses cosmological parameters from Planck \citep{2016A&A...594A..13P}, where $\Omega_M = 0.3089$, $\Omega_{\Lambda} = 0.6911$, $\Omega_b = 0.0486$, and $h = 0.6774$.

\subsection{Black hole growth in the {\sc TNG}}\label{seedingBH_TNG}

{\sc TNG} sows $ 8\times10^{5}\, h^{-1}\, \mathrm{M}_{\odot}$ black hole seeds when halos reach a critical mass of $ 5\times10^{10}\, h^{-1}\, \mathrm{M}_{\odot}$ \citep{2017MNRAS.465.3291W}. After being seeded, black holes grow purely through Bondi accretion limited by the Eddington rate or by merging with another black holes while galaxies merge. {\sc TNG} assumes that black holes accrete at the Bondi rate, but limited by the Eddington rate. This accretion rate is obtained by taking the minimum of both rates, $\dot{m}_{\mathrm{Bondi}}$ (equation \ref{eq:BH_threshold}).

\begin{equation}
\dot{m}_{\mathrm{Bondi}}= \frac{4\uppi G^2 m_{\mathrm{BH}}^2 \rho}{c_{s}^3},\label{eq:BH_Bondi}
\end{equation}

\begin{equation}
\dot{m}_{\mathrm{Edd}}= \frac{4\uppi G m_{\mathrm{BH}} m_p}{\epsilon_r \sigma_T} \ c,\label{eq:BH_Edd}
\end{equation}

\begin{equation}
\dot{m}_{\mathrm{BH}}= {\rm min}(\dot{m}_{\mathrm{Bondi}},\,\dot{m}_{\mathrm{Edd}}),\label{eq:BH_Mdot}
\end{equation}

\noindent
where $c_{s}$ is the sound speed, $\rho$ is the local hot gas density,
$\epsilon_r$ is the radiative accretion efficiency, and $\sigma_T$ is the Thompson cross-section.

When running {\sc TNG} using black hole seeds below $ 10^{6}\, \mathrm{M}_{\odot}$, the black holes take more than a Gyr to grow to supermassive black hole (SMBH) size because the Bondi accretion rate scales as $\dot{m}_{Bondi} \propto m_\mathrm{BH}^{2}$ \citep{2017MNRAS.465.3291W}. Therefore, black hole growth gets delayed as more gas is needed in the center of halos for rapid growth. This means that black holes in $ 10^{12}\, \mathrm{M}_{\odot}$ halos have not ended their rapid accretion period at redshift 0. In turn, there is less feedback energy injected into galaxies, which causes significant cold gas accumulation. As a result, stars form at higher rates, growing galaxies with higher stellar masses over the whole range of halos. Additionally, changing the threshold of the halo mass by which a black hole should be seeded has similar results. Lowering the black hole seed mass while increasing the halo mass threshold requirement to seed the black hole delays black hole growth although not as significantly as in the previous described case \citep{2017MNRAS.465.3291W}.

\subsection{AGN feedback in the {\sc TNG}}\label{AGNfeedbackTNG}

The {\sc TNG} simulation includes gas dynamics, which enables it to model the effects of feedback mechanisms in more detail relative to SAMs. In {\sc TNG}, AGN feedback energy is injected into gas cells directly, and its effects on the gas are then tracked. {\sc TNG} uses two distinct feedback modes: the \textit{thermal mode} at high-accretion rates and the \textit{kinetic mode} at low-accretion rates \citep{2017MNRAS.465.3291W}. The \textit{thermal mode} introduces thermal energy isotropically to the surrounding gas. The \textit{kinetic mode} injects directed momentum kicks into the local environment. The AGN feedback model in {\sc Dark Sage} also has two modes (as described in section \ref{AGNfeedback_DS}), but they operate differently in terms of physics and when these modes are active. The \textit{quasar mode}, most closely analogous to the \textit{thermal mode} in {\sc TNG}, operates during mergers and instabilities as they couple to the ISM. The \textit{radio mode}, similar to the \textit{kinetic mode} in {\sc TNG}, operates at all times, becoming more important at higher stellar masses. The \textit{kinetic mode} in {\sc TNG}, by construction, is particularly effective in high-mass galaxies \citep{Pillepich2018FirstGalaxies, Springel2018FirstClustering}, while the \textit{thermal mode}, on the other hand, is more effective in low-mass galaxies \citep{2019MNRAS.490.3234N}. As a result, it is easier to heat and disperse the gas.

{\sc TNG} uses a combination of the Bondi--Hoyle (equation \ref{eq:BH_Bondi}) and the Eddington (equation \ref{eq:BH_Edd}) accretion rates to determine if the AGN feedback should be \textit{thermal mode} or \textit{kinetic mode} \citep{2017MNRAS.465.3291W}. A black hole mass dependent threshold, $\chi$, is calculated to determine the energy output. 

\begin{equation}
\chi = {\rm min}\left[0.002\left(\frac{m_{\mathrm{BH}}}{10^8 \, \mathrm{M}_{\odot}}\right)^2,\, 0.1 \right],\label{eq:BH_threshold}
\end{equation}

\noindent
If the Eddington ratio $\dot{m}_\mathrm{Bondi}/\dot{m}_\mathrm{Edd}$ exceeds $\chi$, then the \textit{thermal mode} feedback takes place. The energy output becomes:

\begin{equation}
\dot{E}_{\mathrm{thermal \ AGN}} = 5.66 \times 10^{42} \mathrm{erg} \ \mathrm{s}^{-1} \ \frac{\dot{m}_{\mathrm{BH}}}{5 \times 10^{-3}\, \mathrm{M}_{\odot} \ \mathrm{yr}^{-1}},\label{eq:E_thermal}
\end{equation}

\noindent
If instead the Eddington ratio is less than $\chi$, the energy is distributed in kinetic mode at a rate a factor of 10 higher than in ther thermal mode for the same $\dot{m}_\mathrm{BH}$:

\begin{equation}
\dot{E}_{\mathrm{kinetic \ AGN}} = 5.66 \times 10^{43} \mathrm{erg} \ \mathrm{s}^{-1} \ \frac{\dot{m}_{\mathrm{BH}}}{5 \times 10^{-3}\, \mathrm{M}_{\odot} \ \mathrm{yr}^{-1}},\label{eq:E_kinetic}
\end{equation}

\noindent 
At the start, black holes have low accretion rates. After enough black hole mass growth, the feedback strengthens, and the black hole gas supply becomes self-regulated. In {\sc TNG}, the star formation efficiency is highest around halo masses of $10^{12}\, \mathrm{M}_{\odot}$. To quench massive halos, SMBHs transition to the low-accretion state and remain as such throughout their ongoing evolution \citep{2017MNRAS.465.3291W}. This transition takes place around black hole masses of $10^{8}\, \mathrm{M}_{\odot}$ by design. Because of the tight relationship between black hole mass and bulge mass (or stellar mass), the \textit{kinetic mode} feedback takes over for galaxies above stellar masses of $10^{10.25}\, \mathrm{M}_{\odot}$, greatly reducing star formation rates.

The \citet{2017MNRAS.465.3291W} prescription for AGN feedback successfully reproduces a bimodal galaxy color distribution \citep{Nelson2018FirstBimodality}, which results in a UVJ diagram similar to $z\sim 1$ observations \citep{nagaraj2022b} when using an empirical dust model \citep{nagaraj2022a}. The updated AGN feedback also results in TNG's substantially-improved hot gas properties compared to the original Illustris \citep{pop2022}. Despite these successes, the TNG black hole feedback model is essentially phenomenological given the fundamental uncertainty in the physical drivers and the vast difference in scale between the resolution of the simulation and physical scales like the Bondi radius \citep{2015ARA&A..53...51S}. 

\section{Stellar Mass-to-Halo mass relation for {\sc TNG300} and {\sc Dark Sage} galaxies}\label{SMHMR_section}

\begin{figure}[hbt!]
\centering
\includegraphics[width=\columnwidth]{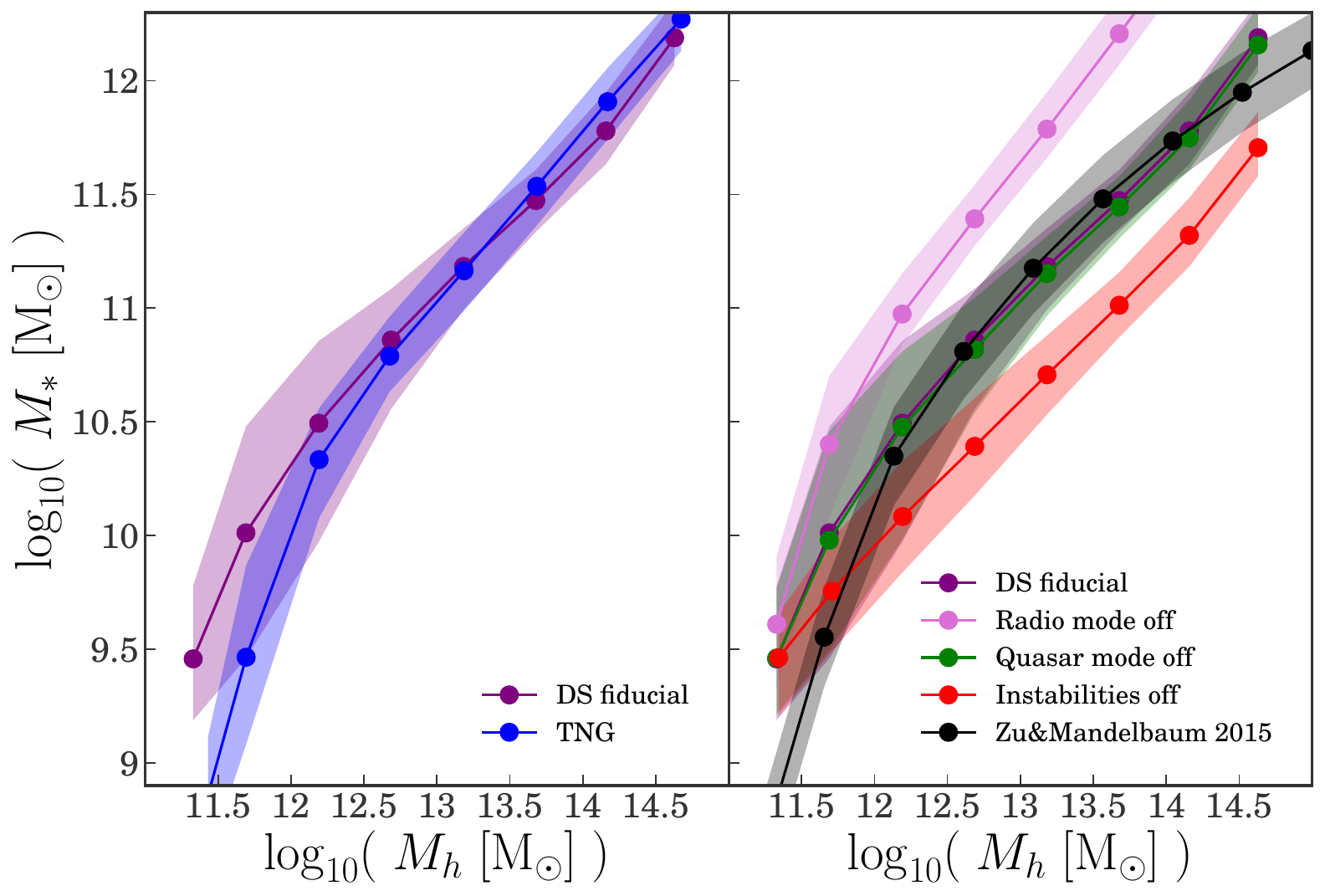}
\caption{Stellar mass as a function of halo mass for central galaxies using: {\sc Dark Sage} fiducial (purple), {\sc TNG} (blue) (left panel), and {\sc Dark Sage} when we turn off disk instabilities, \textit{radio}, and \textit{quasar} mode feedback (red, violet and green, respectively on the right panel). The black line on the right panel shows \citet{Zu2015MappingDR7}'s SMHM empirical relation taken from observational constraints from galaxy-galaxy lensing and galaxy clustering. All solid lines connecting the dots show the median value binned by halo mass. The shaded region enclose 68 percent of the data. Above the quenching mass, {\sc Dark Sage} fiducial galaxies traces {\sc TNG}'s SMHMR. Below that mass, {\sc Dark Sage} overproduces stellar mass galaxies.
Surprisingly, when we turn off \textit{radio mode} feedback and instabilities, the scatter in stellar mass at fixed halo mass reduces significantly at all halo masses. Additionally, all models predict different slopes.}\label{fig:SMHMR_plot}
\end{figure}

Figure \ref{fig:SMHMR_plot} shows the stellar mass-halo mass relation for central galaxies using the fiducial and various modifications of the semi-analytic model {\sc Dark Sage}. We compare these with {\sc TNG} and observational constraints from \citep{Zu2015MappingDR7}. 
For halo masses above $10^{12.5}\, \mathrm{M}_{\odot}$, {\sc Dark Sage} fiducial traces {\sc TNG} galaxies farily well \footnote{Our stellar masses include a fraction of what {\sc Dark Sage} by default assigns to the Intracluster Mass component; see Appendix \ref{app:intracluster_stars_a} for details.} (left panel). Below this mass, {\sc Dark Sage} tends to overestimate stellar masses, and the scatter increases significantly. It is important to note that {\sc TNG300} was calibrated using the same model parameterizations as the {\sc TNG100} \citep{2019ComAC...6....2N}, so it is known to overproduce high stellar mass galaxies. In this work, we are not including the ``rescaling factor" used to scale the particle masses and resolution of the simulation \citep{Pillepich2018FirstGalaxies}. While {\sc TNG} galaxies have an approximately constant scatter, {\sc Dark Sage} has a significantly larger scatter at low halo masses that decreases with increasing halo mass. 

The right hand side panel shows that, surprisingly when the \textit{radio mode} feedback or disk instabilities are turned off in {\sc Dark Sage}, the scatter in stellar mass is significantly reduced, closely resembling that of {\sc TNG}. Interestingly, a similar experiment only turning off \textit{quasar mode} shows no significant change in the stellar-mass scatter nor SMHMR slope. This simple experiment shows that \textit{radio mode} feedback and disk instabilities contribute significantly to the large scatter in stellar mass at fixed halo mass. Because \textit{radio mode} feedback dumps energy into the hot halo to prevent hot gas from cooling, we see that when it is turned off, this hot gas cools rapidly to form stars and yielding much larger stellar masses at any given halo mass. 

Turning off disk instabilities leads to lower stellar masses at every halo mass. This happens because disk instabilities account for the majority of the star formation in the model (see fig. 1 of \citet{10.1093/mnras/stx1596}). {\sc Dark Sage} uses three star formation channels: merger-driven starbursts, H$_2$ gas content, and disk-instabilities. Without disk instabilities, gas is stranded at large radii where the longer dynamical times and lower metallicities slow the rate of star formation \citep{Stevens2016}. Additionally, turning off instabilities reduces the growth of black holes, which in turn reduces the effect of \textit{radio-mode} feedback in galaxies.

\begin{figure*}[t]
\centering
\includegraphics[width=2.0\columnwidth, clip]{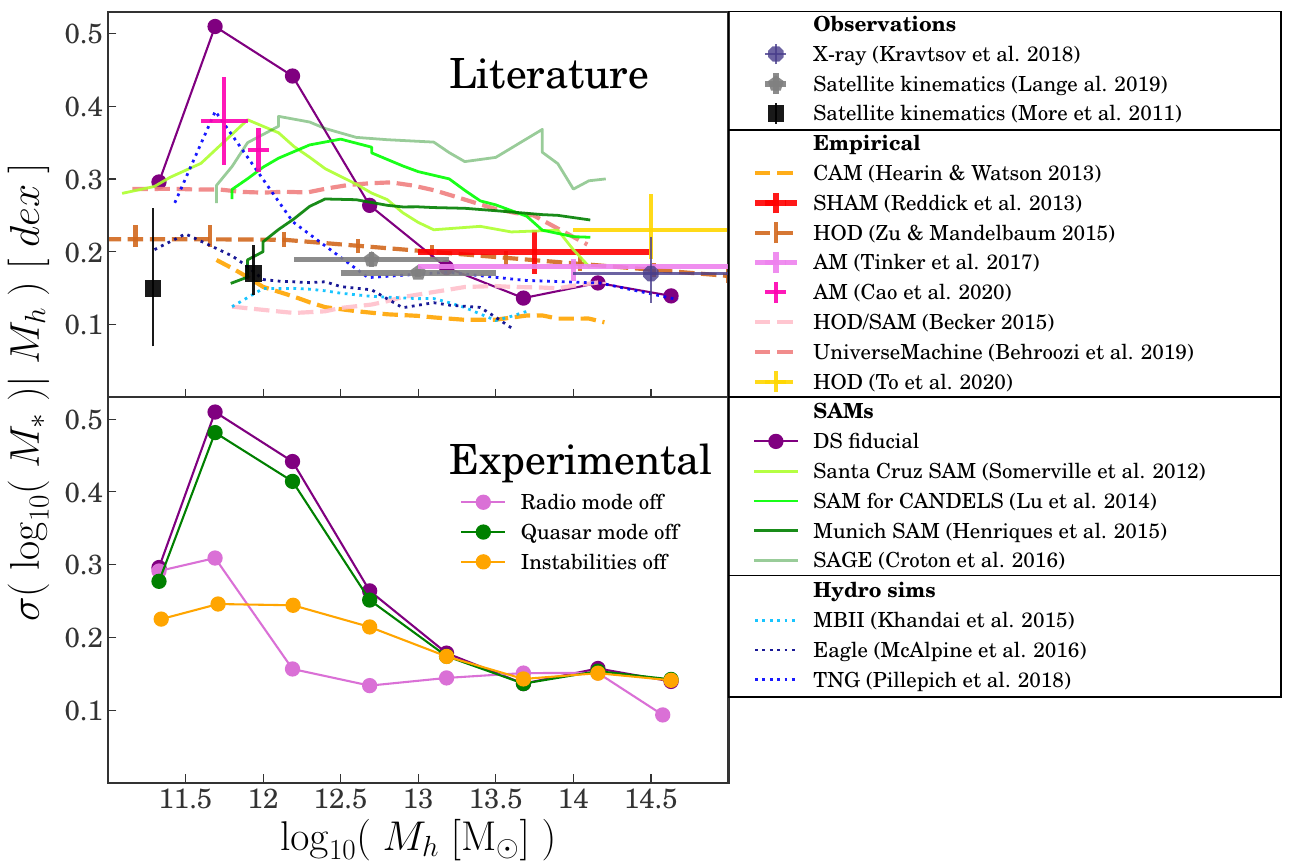}
\caption{The stellar mass scatter -- computed as half the 16th--84th interpercentile range -- at fixed halo mass $(\sigma(\log_{10}(M_* | M_h)))$ as a function of halo mass for central galaxies. The top panel shows literature results from: a) observations of X-ray measurements shown in dark slate blue octagons \citep{Kravtsov2018StellarHalos}, satellite kinematics shown in grey pentagons \citep{2019MNRAS.487.3112L} and black squares \citep{2011MNRAS.410..210M}. b) empirical models from \citet{Hearin2013TheColour}, \citet{2013ApJ...771...30R}, \citet{Zu2015MappingDR7}, \citet{2017ApJ...839..121T}, \citet{2020MNRAS.498.5080C}, \citet{2015arXiv150703605B}, \citet{Behroozi2019Universemachine:010}, and \citet{2020ApJ...897...15T} are shown as dashed lines using varying light colors, c) SAMs from \citet{2012MNRAS.423.1992S}, 
 \citet{2014ApJ...795..123L}, \citet{Henriques2015}, and \citet{ Croton2016} are shown as solid lines with varying green shades, and d) cosmological hydrodynamical simulations from \citet{2015MNRAS.450.1349K}, \citet{2016A&C....15...72M}, and \citet{Pillepich2018FirstGalaxies} are shown as dotted lines with different blue shades. Several lines are taken from both \citet{Wechsler2018} and \citet{2020MNRAS.498.5080C}. Note that for \citet{2013ApJ...771...30R}, \citet{2017ApJ...839..121T}, \citet{2020MNRAS.498.5080C}, \citet{2019MNRAS.487.3112L}, and \citet{2020ApJ...897...15T}, the extent displayed in the x-error bar is the range of halo masses estimated to be probed with each study. Both top and bottom panels include the same {\sc Dark Sage} fiducial line in purple. The bottom panel shows the experimental modifications to {\sc Dark Sage} when we turn off disk instabilities (orange), \textit{quasar} (green), and \textit{radio} (pink) mode feedback. Below the quenching mass, the stellar mass scatter is significantly larger in {\sc Dark Sage} fiducial compared to {\sc TNG} and most models. We find that when we turn off \textit{radio mode} feedback, the scatter in stellar mass at fixed halo mass reduces significantly at all halo masses.}\label{fig:SMHMRscatter_plot}
\end{figure*}

Figure \ref{fig:SMHMRscatter_plot} shows the stellar mass scatter -- computed as half the 16th--84th interpercentile range -- at fixed halo mass $\sigma(\log_{10}(M_* | M_h))$ as a function of halo mass. Below the quenching masses, {\sc Dark Sage} fiducial galaxies have a higher scatter in stellar mass than in observations, {\sc TNG}, and roughly all other models. Unlike {\sc Dark Sage} fiducial, the scatter in {\sc TNG} is generally constant at $\sim 0.15\ \mathrm{dex}$, except at low halo masses where the scatter begins to approach $0.3\ \mathrm{dex}$. This low scatter prediction is consistent with other hydrodynamical simulations \citep{2015MNRAS.450.1349K, 2016A&C....15...72M, Pillepich2018FirstGalaxies}, empirical models \citep{Hearin2013TheColour, 2013ApJ...771...30R, 2015arXiv150703605B, 2017ApJ...839..121T, Behroozi2019Universemachine:010, 2020MNRAS.498.5080C, 2020ApJ...897...15T}, and observations from satellite kinematics \citep{2011MNRAS.410..210M, 2019MNRAS.487.3112L}\footnote{\citet{2020ApJ...897...15T} and \citet{2019MNRAS.487.3112L} present values of the scatter in luminosity. Here, we assume that the scatter in luminosity is equal to the stellar mass scatter. To get halo mass in \citet{2019MNRAS.487.3112L}, we take the luminosity bins that has the most secondaries for the blue and red galaxies given in their table 1. Then, we use the top left plot on their figure 4 to roughly estimate the halo mass within the luminosity bin. We caution the readers in interpreting these points.}, galaxy-galaxy lensing \citep{Zu2015MappingDR7} and X-ray \citep{Kravtsov2018StellarHalos} measurements. Around the quenching mass, {\sc Dark Sage}'s scatter almost gets as large as $0.6\ \mathrm{dex}$, which is higher than other semi-analytic models. However, when the \textit{radio mode} is turned off in {\sc Dark Sage}, the scatter drops to about $0.2\ \mathrm{dex}$. We see that turning off \textit{quasar mode} feedback does not reduce the scatter significantly, but interestingly, without disk instabilities, the stellar-mass scatter drops just below $0.25\ \mathrm{dex}$ and keeps constant as a function of halo mass similar to other SAM predictions. 

From the perspective of galaxy formation models, \citet{Wechsler2018} compares $\sigma(\log_{10}(M_* | M_h))$ predictions from empirical, SAMs, and hydrodynamical simulations. Several \textit{empirical models} and hydrodynamical simulations agree that the scatter in stellar mass at fixed halo mass is constrained to be $\la 0.2\ \mathrm{dex}$ for galaxies with halo masses above $10^{12}\, \mathrm{M}_{\odot}$. However, SAMs predict a much larger scatter in stellar mass, ranging between $0.2-0.4\ \mathrm{dex}$ at halo masses between $10^{11}-10^{15}\, \mathrm{M}_{\odot}$. These predictions are in line with the \citet{Behroozi2019Universemachine:010} \textit{empirical model}, which uses star formation histories that are matched to the halo formation time.

There is further work to be done to help constrain the scatter in stellar mass at low halo masses ($M_h < 10^{11}\,{\rm M}_{\odot}$). It is possible that the increased stellar mass scatter in {\sc TNG} is due to the unrealistic blow-outs of gas just below the quenching mass, which leads to overly massive galaxies and low stellar metallicities in their central regions. This is only known to happen for halos with underdense gas \citep{2019ComAC...6....2N}. Another possibility is the resolution of {\sc TNG300}. Figure 8 in \citet{Wechsler2018} shows the scatter in stellar mass at fixed halo mass for {\sc TNG300} and {\sc TNG100}. They show that at fixed halo mass for $M_h < 10^{12}\,{\rm M}_{\odot}$, {\sc TNG300} galaxies have a larger stellar mass scatter. The scatter difference between the simulations increases up to $\sim 0.1 \ \mathrm{dex}$ with decreasing halo mass. Using {\sc Marvel-ous Dwarfs} \citep{2019ApJ...874...40M} and {\sc DC Justice League} \citep{2021ApJ...906...96A} zoom-in cosmological simulations, \citet{Munshi_2021} finds that galaxies between $M_* \sim 10^{6-9}\,{\rm M}_{\odot}$ and $M_h \sim 10^{8-11}\,{\rm M}_{\odot}$ have a constant scatter in stellar mass of $0.3 \ \mathrm{dex}$. In the case of empirical models, at the low halo mass regime, the standard techniques that match galaxies with dark matter halos from cosmological simulations would not be as reliable since the assumption of a nonlinear monotonic relationship between the stellar mass and the halo mass may break. This is important given that there is an explicit degeneracy between the observationally-inferred scatter and the SMHM slope \citep{2017MNRAS.464.3108G}. An exploration of the scatter in stellar mass at $M_h < 10^{12}\,{\rm M}_{\odot}$ will follow in future works. To understand the stellar-mass scatter in {\sc Dark Sage}, we look at the galaxy mass trajectory alongside the SMHMR within a time frame.

\begin{figure}[t]
\centering
\includegraphics[width=\columnwidth, clip]{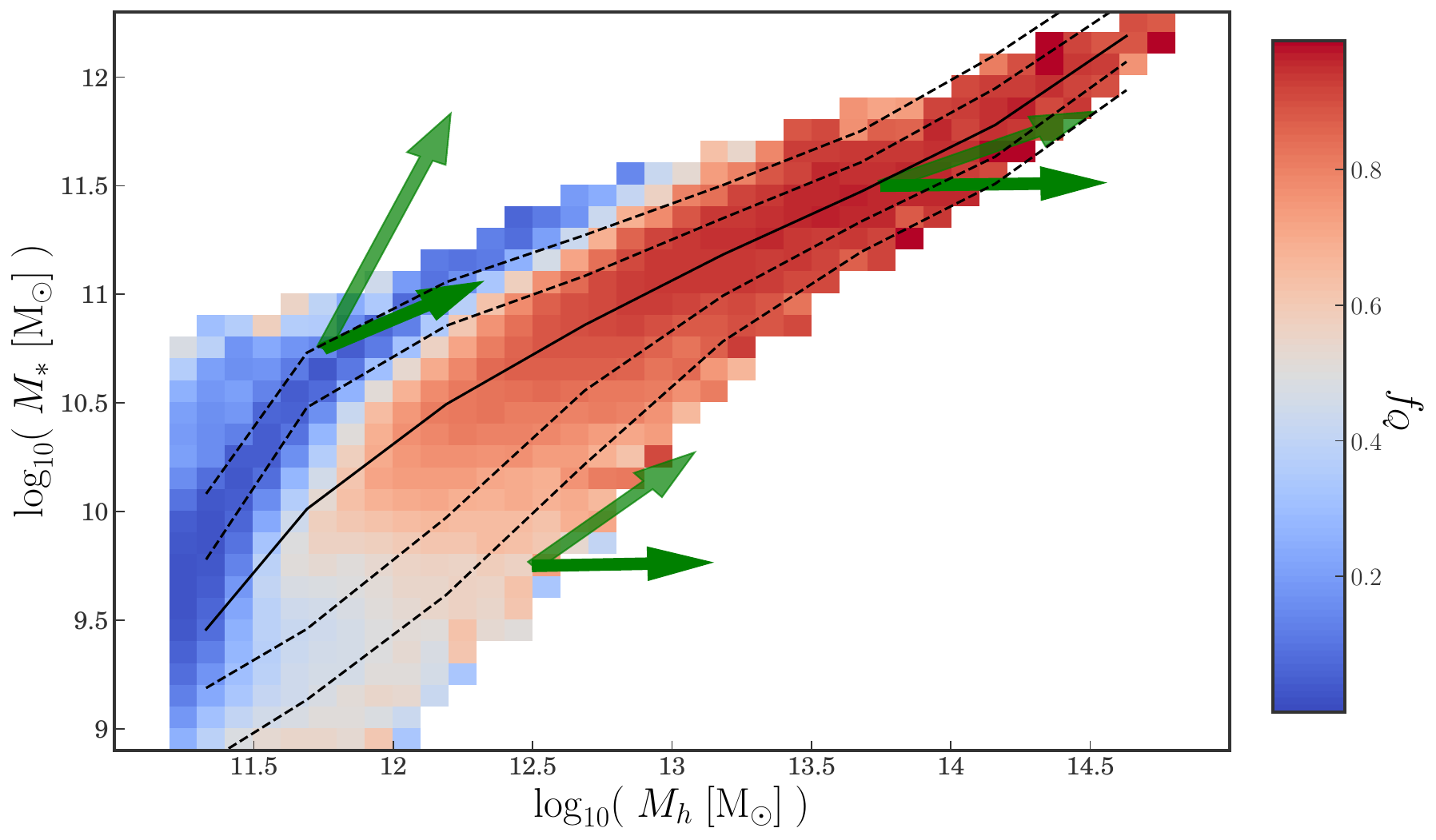}
\caption{The stellar-to-halo mass relation colored by the quenched fraction ($\log_{10} \left( \mathrm{sSFR} \ \left[ \mathrm{yr}^{-1} \right] \right) < -11.0$) using {\sc Dark Sage} fiducial. The black solid line shows the median binned by halo mass. The dashed lines enclose 68 and 95 percent of the data, respectively. The green arrows show the average mass growth of halos and stellar mass in three different (0.5 dex)$^2$ bins in the stellar mass-to-halo mass space\footnote{The arrows' x- and y- extents are $\Delta t\ d\log_{10} M_h/dt$ and $\Delta t\ d\log_{10} M_*/dt$ respectively, where we have chosen $\Delta t = 10/\ln(10)$ Gyr. For simplicity the halo growth rate is taken from the fitting formula in \citet{bouche2010} and the stellar mass growth is assumed to come from star formation only.}. Within each bin we sort galaxies according to the ratio of their specific star formation rate to the specific halo mass growth rate, i.e. (the arctangent of) the angle the arrow would make with the x-axis, and plot the average arrow derived from the upper and lower halves of the distribution. Galaxies in the upper and lower regions tend to increase the stellar mass scatter.}
\label{fig:growthmass_heatmap_plot}
\end{figure}

Figure \ref{fig:growthmass_heatmap_plot} shows the stellar-to-halo mass relation colored by the quenched fraction for {\sc Dark Sage} fiducial galaxies. The top-left region and the bottom region, namely the regions with a halo mass around $10^{12}\, {\rm M}_\odot$, have a substantial fraction of galaxies evolving along the existing SMHMR in the simulation, and a substantial fraction moving away from the existing relation. These regions would therefore tend to increase the scatter in the SMHMR, all else equal. Investigating the galaxies in these regions further, we have found what we believe to be the principle cause of the large scatter in the fiducial {\sc Dark Sage} run: the existence of low-mass black holes in high-stellar mass galaxies, and the existence of high-mass black holes in low-stellar mass galaxies for $M_h \sim  10^{12}\, {\rm M}_\odot$ (see Section \ref{fiducial_AGNfeedback}). The high-mass region whose trajectories are shown in Figure \ref{fig:growthmass_heatmap_plot} shows galaxies predominantly moving in the same, rightward, direction, though of course caution must be warranted in interpreting these arrows, since much of the stellar mass growth in such systems is expected to come from mergers \citep{RodriguezGomez2016}. Given the averaging out of deviations from some mean relation that generically occurs in mergers, we expect halos in this regime will steadily experience decreasing scatter in stellar mass. We now explore ways to reduce the scatter in stellar mass taking into account the significant contribution of the black hole mass and feedback.

\section{Experiments in modifying Dark Sage's SMBH Population and their feedback}\label{BHM_seedingAGN}

\begin{figure*}[hbt!]
\centering
\includegraphics[width=1.55\columnwidth]{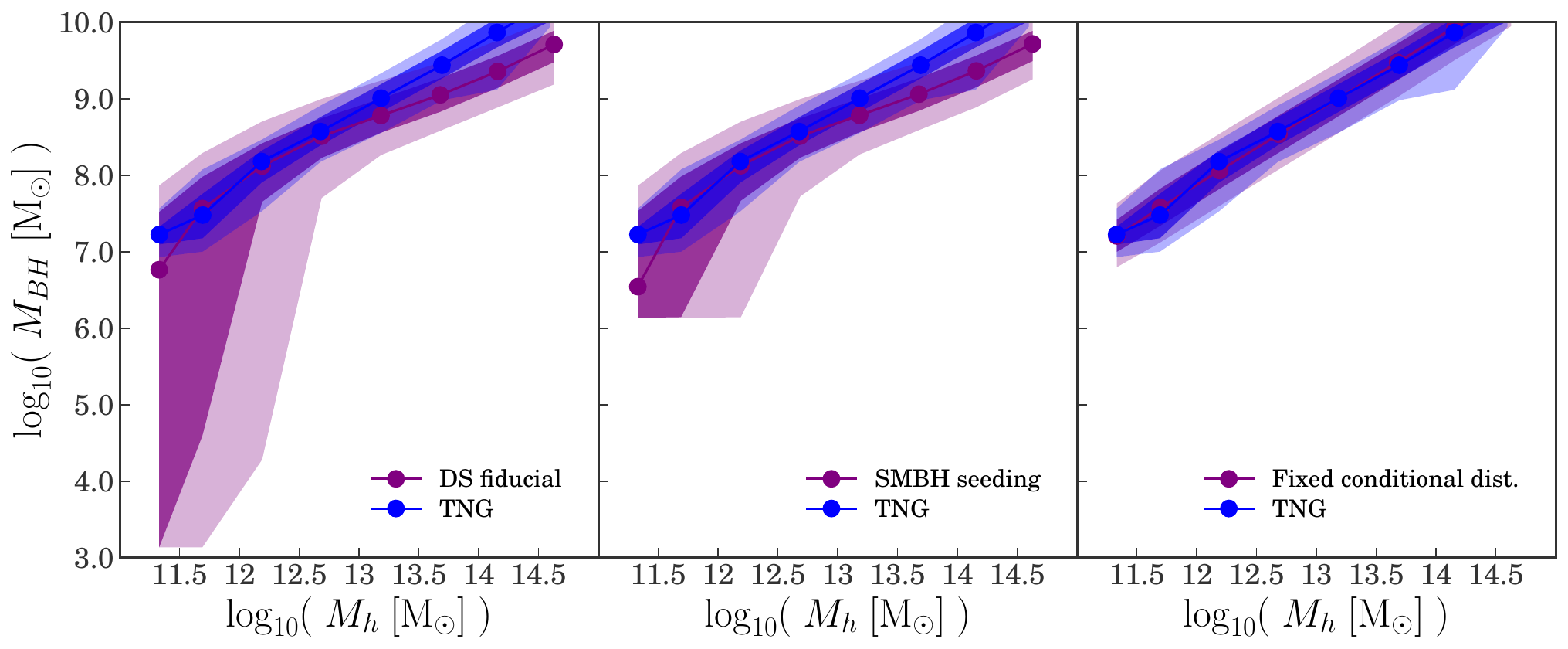}
\caption{Black hole mass as a function of halo mass for {\sc Dark Sage} (purple) and {\sc TNG} (blue) galaxies. In all panels, the blue line is unchanged. The left panel shows fiducial {\sc Dark Sage}. The middle panel shows {\sc Dark Sage} results when the model seeds black holes like in {\sc TNG} (i.e. if a halo of $5 \! \times \! 10^{10} \, h^{-1}\, \mathrm{M}_{\odot}$ does not have a black hole of $10^{6}\, \mathrm{M}_{\odot}$, then assign one). The right panel shows {\sc Dark Sage} galaxies when we apply our fixed conditional distribution, where we force galaxies to follow the black hole mass--halo mass median relation shown in {\sc TNG} including the following black hole mass scatter values: 0, 0.2, and 0.4$\ \mathrm{dex}$. In this plot, we only display the black hole mass--halo mass median relation with a black hole mass scatter of 0.2$\ \mathrm{dex}$.}\label{fig:BHM_Mh_BHseeding_plot}
\end{figure*}

We begin by exploring how black hole formation changes the scatter in stellar mass at fixed halo mass. We present two cases as shown in Figure \ref{fig:BHM_Mh_BHseeding_plot}: the SMBH seeding and the fixed conditional distribution. For the SMBH seeding case, we allow {\sc Dark Sage} galaxies to grow their black holes as they normally do, through gas accretion (see equation \ref{eq:BH_Bondi}) and mergers (equation \ref{eq:BH_quasar}) as described in section \ref{seedingBH_DS}. However, in this model, we now seed a black hole mass of $10^{6}\, \mathrm{M}_{\odot}$ for every halo that reaches $5 \! \times \! 10^{10} \, h^{-1} \, \mathrm{M}_{\odot}$. {\sc Dark Sage} initializes black holes with zero mass in the fiducial run. Here, we are basically mimicking how {\sc TNG} treats black hole seeding as described in Section \ref{seedingBH_TNG}. 

In the case of the fixed conditional distribution model, we remove any form of black hole growth that fiducial {\sc Dark Sage} implements. Instead, we force all black hole masses to follow a pre-ordained conditional distribution at fixed halo mass, centered on the median black hole mass--halo mass relation from {\sc TNG}. {\sc Dark Sage} now assigns black hole mass using equation \ref{eq:MBHMh_model}.

\begin{equation}
\frac{m_{\mathrm{BH}}}{10^{10}\, \mathrm{M}_{\odot}} = 
\left( \frac{m_{\mathrm{halo}}}{10^{10}\, \mathrm{M}_{\odot}} \right)^{0.95} \ \times 10^{-4.04+x},\label{eq:MBHMh_model}
\end{equation}

\noindent 
To control the black hole mass--halo mass scatter, we add $x$, a random variable normally distributed with mean zero and standard deviation $\sigma_\mathrm{BH}$. Each galaxy's value of $x$ remains fixed over time along the main progenitor history.

It is important to note that when we impose this distribution of black hole masses in {\sc Dark Sage} at all times over the course of the simulation, we are decoupling some of the usual relationships between black hole mass and cumulative energy injection by AGN feedback. Of these the most consequential is the typical relationship between $\dot{m}_\mathrm{heat} \propto L_\mathrm{BH,R}$ and $\dot{m}_\mathrm{Bondi,R}$ in radio mode feedback. Over any time period where radio mode feedback is dominant, the fiducial setup of {\sc Dark Sage} would yield an energy imparted into the CGM
\begin{equation}
\label{eq:coupling}
    \Delta E = \int \eta\, c^2\, \dot{m}_\mathrm{BH}\, dt = \eta\, c^2\, \Delta m_\mathrm{BH},
\end{equation}
where $\Delta A$ denotes the difference in quantity $A$ between the start and end of the time period. While the first equality in Equation \ref{eq:coupling} holds, the second no longer does when we impose a given conditional black hole mass distribution because $\Delta m_\mathrm{BH} = 10^{5.96+x}\Delta (m_\mathrm{halo}/10^{10}\, {\rm M}_\odot)^{0.95}\, {\rm M}_\odot$, and not $\int \dot{m}_\mathrm{BH}\, dt$. Effectively the ratio of $\Delta E$ to $\Delta m_\mathrm{BH}$ goes from being a constant to a nonlinear function of halo mass and the random variable $x$. This disconnect can be particularly non-intuitive for large values of $x$, which produce large values of $\Delta m_\mathrm{BH}$ without correspondingly large values of $\Delta E$. This approach has allowed us to quickly explore the effects of changing the distribution of black hole masses with and without modifying the rate of energy produced by a given black hole. One could imagine an alternative approach involving exploring the parameter space of the SAM to attempt to reproduce something close to the black hole mass--halo mass relation in TNG while leaving the coupling in Equation \ref{eq:coupling} intact, but this is of course a much more expensive process and would not allow us to easily control $\sigma_\mathrm{BH}$ as we do here.

Figure \ref{fig:BHM_Mh_BHseeding_plot} shows the black hole mass - halo mass relation for both the SMBH seeding and fixed conditional distribution models. We see that changing black hole seeding in our model closely reproduces the black hole mass - halo mass relation seen in {\sc TNG}. Because star formation rate and quenching of a galaxy are strongly tied to black hole feedback, we now turn to dissecting the effect of black hole mass within three different AGN feedback models: {\sc Dark Sage} instantaenous AGN feedback, turning off AGN feedback while also turning off cooling for galaxies with black hole masses above $10^{8}\, \mathrm{M}_{\odot}$, and lastly, doing the former while also removing all the cold gas in the interstellar medium. We specifically choose to change both the black hole seeding and AGN feedback in {\sc Dark Sage} in order to reproduce the black hole physics and stellar mass scatter seen in {\sc TNG}. In the following sections, we analyze how both black hole mass and these three modes of feedback contribute to reducing the scatter in stellar mass.

\subsection{Instantaneous AGN feedback model} \label{fiducial_AGNfeedback}

\begin{figure*}[hbt!]
\centering
\includegraphics[width=1.8\columnwidth, clip]{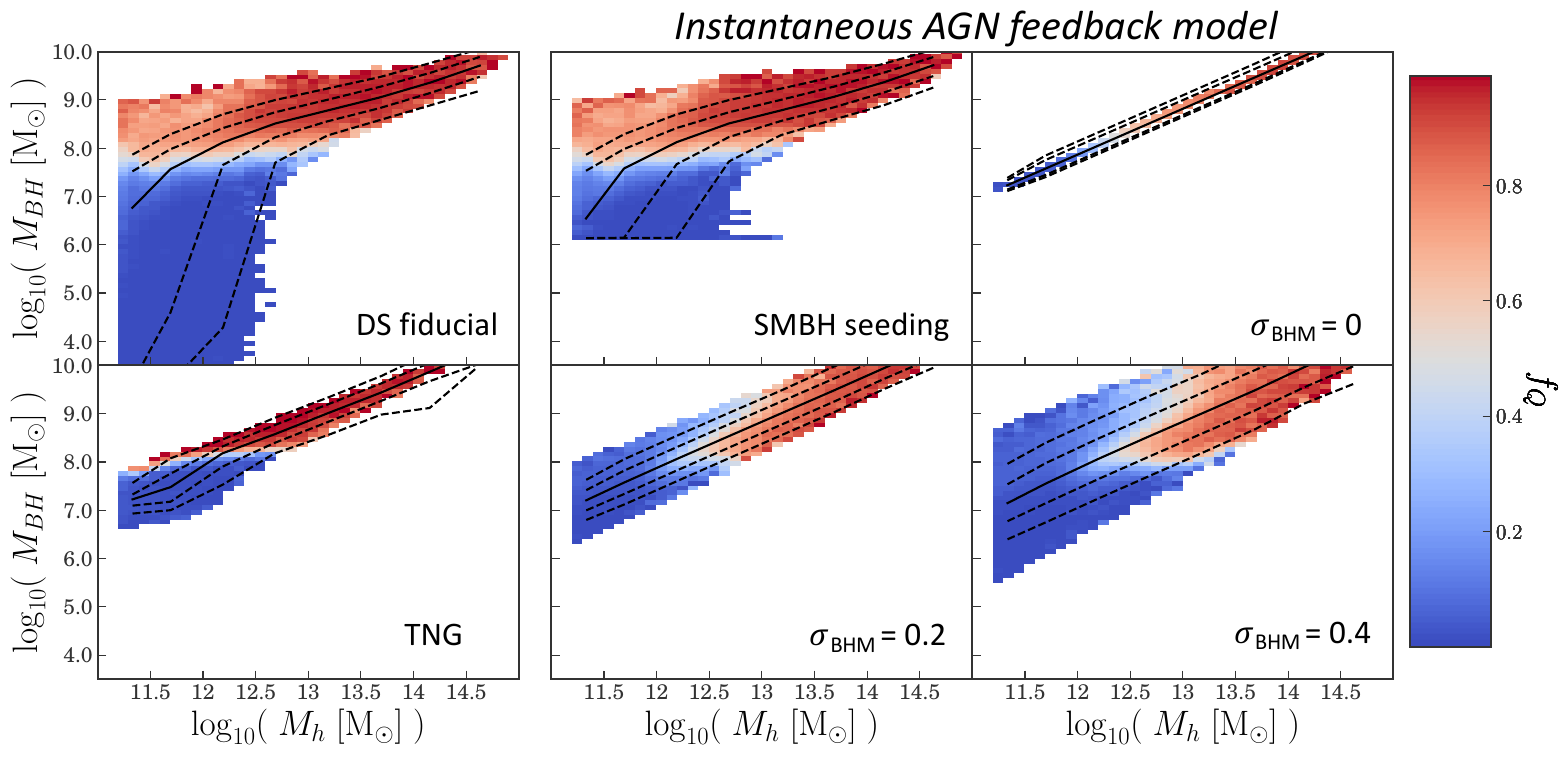}
\caption{Instantaneous AGN feedback model: Black hole mass as a function of halo mass colored by the quenched fraction ($\log_{10} \left( \mathrm{sSFR} \ \left[ \mathrm{yr}^{-1} \right] \right) < -11.0$) for {\sc Dark Sage} fiducial (top left), {\sc TNG} (bottom left), SMBH seeding and several sigma runs with a fixed conditional distribution. For the fixed conditional distribution case, we assume a scatter of 0, 0.2, and 0.4 $\ \mathrm{dex}$. In the case of SMBH seeding, the quenching transition happens around a black hole mass of $10^{8}\, \mathrm{M}_{\odot}$, like in {\sc Dark Sage} fiducial and {\sc TNG}. In the case of the fixed conditional distribution model, both black hole mass and halo mass contribute to suppressing star formation.}
\label{fig:logMvirlogBHM_classicradiomode_plot}
\end{figure*}

In this case, we allow {\sc Dark Sage} to prescribe the \textit{radio mode} and \textit{quasar mode} feedback models as described in section \ref{AGNfeedback_DS}, meaning that the accretion rates\footnote{For the conditional black hole mass model, these ``accretion rates'' are used to evaluate the feedback rates, but are not used to increase the mass of the black holes because the mass of the black holes is set by Equation \ref{eq:MBHMh_model}.} onto the black hole and their corresponding energy output are calculated as usual. Figure \ref{fig:logMvirlogBHM_classicradiomode_plot} shows the black hole mass--halo mass relation colored by the quenched fraction. 
{\sc Dark Sage} fiducial shows that black holes above $\sim 10^{7.5}\, \mathrm{M}_{\odot}$ are quenching their galaxy, similar to {\sc TNG} results. As stated in section \ref{AGNfeedbackTNG}, {\sc TNG} uses a threshold in black hole mass at $\sim 10^{8}\, \mathrm{M}_{\odot}$ to change from \textit{thermal mode} to \textit{kinetic mode} accretion, which explains the sharp transition from active to passive galaxies shown here. Interestingly, {\sc Dark Sage} fiducial naturally produces a similar quenching transition without having any explicit threshold for a qualitative change in feedback. In the SMBH seeding case, galaxies quench when their black holes reach a mass of $\sim 10^{8}\, \mathrm{M}_{\odot}$, much as in fiducial {\sc Dark Sage} and {\sc TNG}. In these three cases, quenching seems to be determined first and foremost by the mass of the black hole, though we do see that some fraction of DS galaxies with black hole masses as high as $10^9 {\rm M}_\odot$ are not quenched at $z=0$. In the case of the fixed conditional distribution model, the relationship between black hole mass and quenching is much less clean-cut, with a prominent population of high-mass black holes that live in star-forming galaxies with halo masses below $\sim 10^{12.5}\, {\rm M}_\odot$ emerging as $\sigma_\mathrm{BH}$ increases. While these black holes are relatively few in number, as they live in the tails of the distribution (whose central 68\% and 95\% at each halo mass are shown with dashed lines), it is striking that in this case high black hole masses are insufficient to quench galaxies. Because the offset from the mean black hole mass-halo mass relation remains constant for the main progenitor branch history of each galaxy in this model, these galaxies have had higher-than-average black hole masses for a long time. However, because the black hole growth is explicitly set by the halo in this model, rather than by feedback, the cumulative radio mode energy output of these black holes is likely much lower than the same-mass black hole in DS fiducial or TNG. This proposition will be tested further in the subsequent subsections by modifications to the feedback scheme.

\begin{figure*}[hbt!]
\centering
\includegraphics[width=1.8\columnwidth, clip]{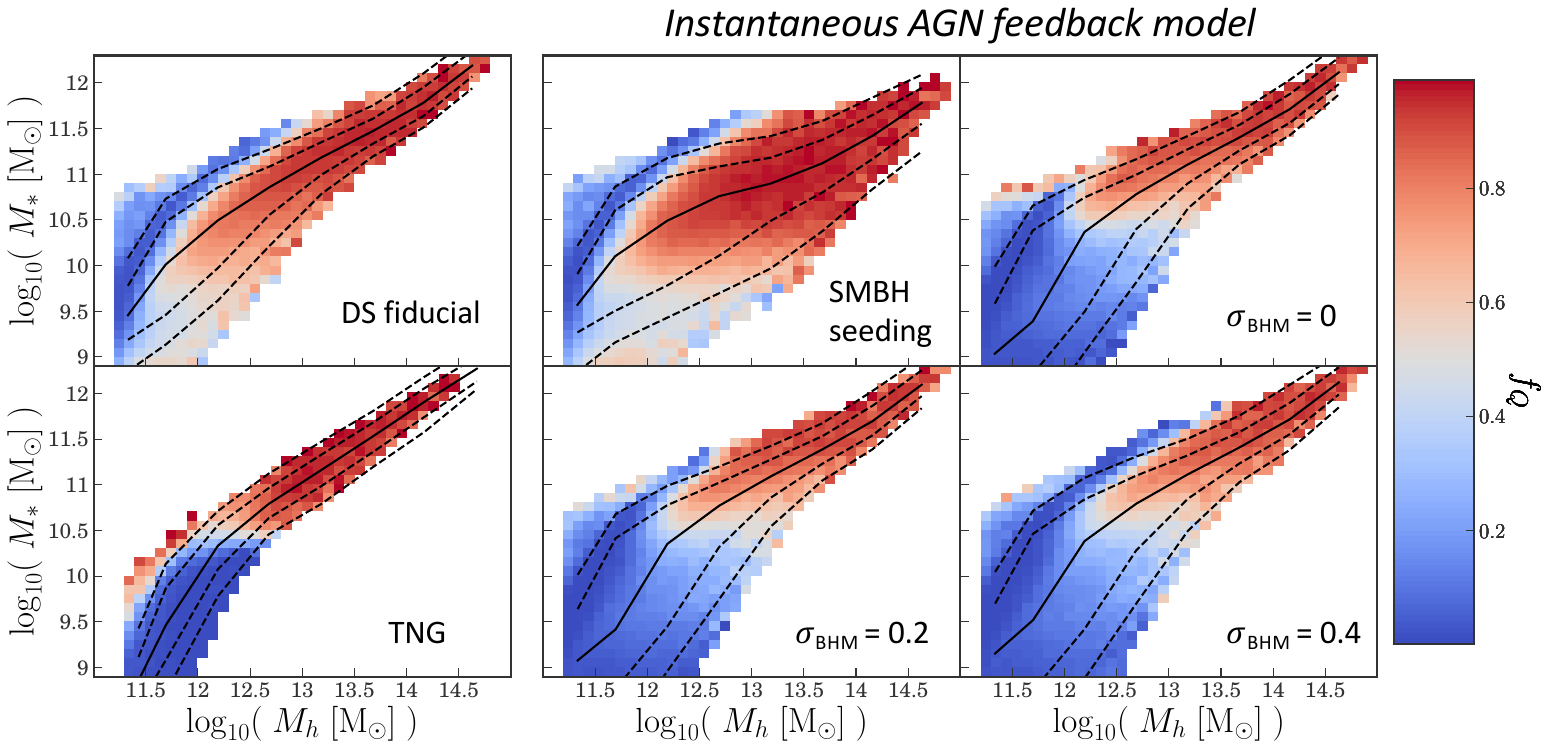}
\caption{Instantaneous AGN feedback model: Stellar-to-halo mass relation colored by the quenched fraction for the same black hole models as in Figure~\ref{fig:logMvirlogBHM_classicradiomode_plot}. We find that the overall scatter in stellar mass is not reduced by the change in black hole seeding for this particular feedback model. We do not see a significant change in the stellar mass scatter when increasing the scatter in black hole mass.}
\label{fig:SMHMR_classicradiomode_plot}
\end{figure*}

These modifications to the distribution of black hole masses are also visible in the SMHMR. Figure \ref{fig:SMHMR_classicradiomode_plot} shows this relation colored by the quenched fraction. The fact that {\sc Dark Sage} seems to quench strongly on halo mass might be a result of black holes being seeded based on halo mass in {\sc Dark Sage}, and how black holes play the main role on galaxy quenching. We see that above stellar mass of $10^{10.25}\, \mathrm{M}_{\odot}$, {\sc TNG} galaxies are quenched. This is likely due to their tight black hole mass-stellar mass relation. As a result, \textit{kinetic mode} black hole accretion, not stellar mass, is likely the primary galaxy quenching mechanism in {\sc TNG}. {\sc Dark Sage} shows a more complex picture. We see a blue strip on the bottom and top left part of the SMHMR, indicating a fair amount of star-forming galaxies. There are also some star-forming galaxies below stellar mass of $10^{10}\, \mathrm{M}_{\odot}$ at all halo masses. This is explained by understanding the distribution of black hole masses in both of these regions. Galaxies that have high stellar masses, but low halo mass have low-mass black holes in their center ($m_{\mathrm{BH}} < 10^{8}\, \mathrm{M}_{\odot}$). Low-mass black holes have their growth stalled, and therefore do not have AGN strong enough to quenched their galaxy, while low-mass quenched galaxies tend to be those with massive black holes.

We find that the overall scatter in stellar mass is not reduced by the change in black hole seeding for this particular feedback model. We do notice an increase in the stellar mass scatter when increasing the scatter in black hole mass. The same star-forming strip in the upper part of the stellar-to-halo mass relation in the fixed conditional distribution models shows up, increasing in halo mass extent with larger values of the black hole mass scatter; the large scatter in black hole mass produces a population of galaxies with halo masses above $10^{12}\, {\rm M}_\odot$ but with black hole masses below $10^8\, {\rm M}_\odot$, which make up this prominent feature.

\begin{figure}[t]
\centering
\includegraphics[width=\columnwidth, clip]{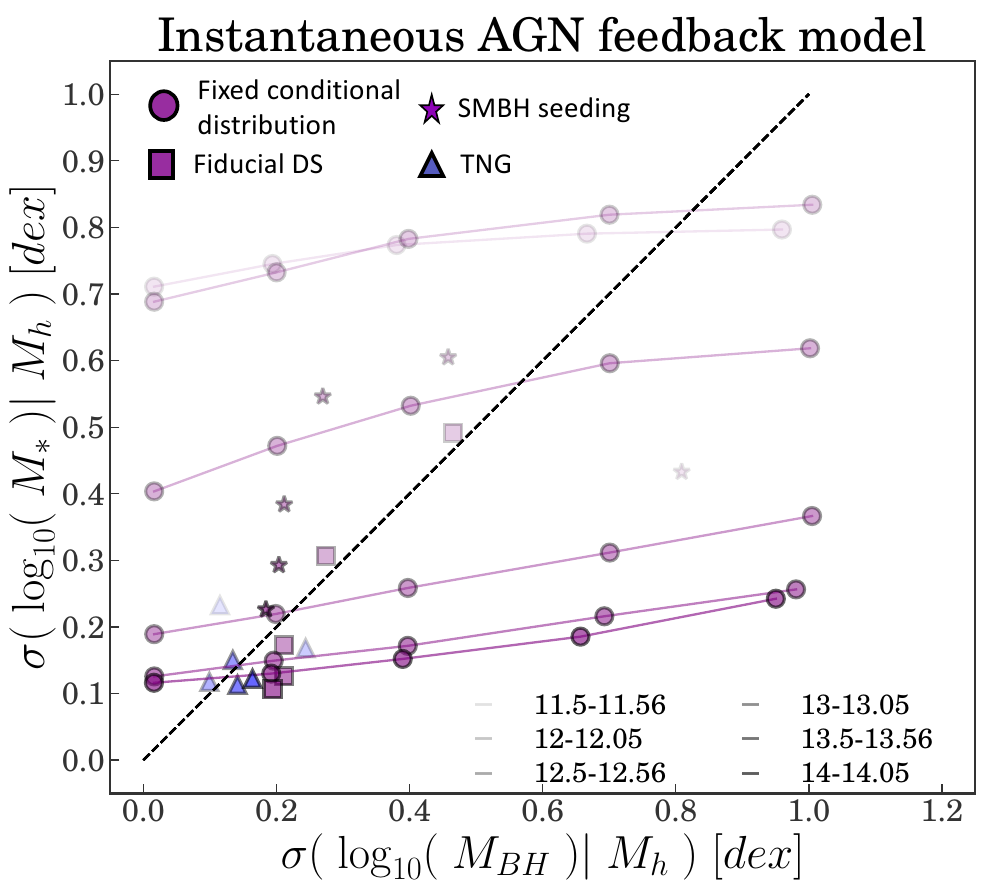}
\caption{The stellar mass scatter at fixed halo mass $\sigma(\log_{10}(M_* | M_h))$ as a function of black hole mass scatter at fixed halo mass $\sigma(\log_{10}(M_{\rm BH} | M_h)$. Here, we adopt the instantaneous AGN feedback model, where the only change we make is the way black hole masses are calculated. Each purple line with dots shows a different halo mass bin. The purple squares and stars show {\sc Dark Sage} fiducial and SMBH seeding data respectively. The blue triangles show {\sc TNG} data. In all data points, the bolder the color, the higher the halo mass bins. The black line is a one-to-one line. When seeding SMBH in {\sc Dark Sage}, there is a slight increase in the scatter in stellar mass at the low mass bins. When implementing the conditional black hole mass model, the scatter in stellar mass worsens for the lowest halo mass bins, when if we assume a zero black hole mass scatter.}
\label{fig:scatterMstarMh_scatterMBHMh_classicradiomode_plot}
\end{figure}

Since we can easily change the scatter in the black hole mass for the conditional black hole mass model, we now show how systematically varying that scatter may affect the scatter in stellar mass. Figure \ref{fig:scatterMstarMh_scatterMBHMh_classicradiomode_plot} presents the stellar mass scatter at fixed halo mass $\sigma(\log_{10}(M_* | M_h))$, in dex, as a function of black hole mass scatter\footnote{The black hole mass scatters are measured from the simulation and can therefore be slightly different from the input value of $\sigma_\mathrm{BH}$. The measured values are slightly smaller than $\sigma_\mathrm{BH}$ especially for the largest values of $\sigma_\mathrm{BH}$ and the highest halo mass bins. This is because in mergers the $x$ value of the primary is retained, and the value of $x$ has some bearing on the stellar mass and hence which galaxy ends up being the primary.} at fixed halo mass $\sigma(\log_{10}(M_{\rm BH} | M_h))$ in several narrow halo mass bins. 

For the fixed conditional distribution model, we find that increasing the scatter in black hole mass systematically increases the scatter in the stellar mass for halos below the quenching mass. The largest effects (slopes) are for the lowest halo mass bins. The SMBH seeding case presents a slightly larger stellar mass scatter compared to fiducial {\sc Dark Sage} in the low halo mass bins. This reinforces a lesson from Figure \ref{fig:SMHMR_classicradiomode_plot}, namely that the difference in seeding tried here has some effect on the SMHMR scatter. We also note that the fiducial {\sc Dark Sage} and SMBH seeding points lie close to the curves for their respective halo masses for halo masses above $\sim 10^{13}\, {\rm M}_\odot$, but for lower masses, the conditional black hole mass distribution experiment produces a scatter in stellar mass far larger than the values given by fiducial {\sc Dark Sage}. These values go up to $0.7\ \mathrm{dex}$ even when we assume a scatter in black hole mass of $0\ \mathrm{dex}$. Lastly, we see how {\sc TNG} galaxies compare with all models, yielding similar results to fiducial {\sc Dark Sage} and our models at the high mass bins. These results show that modifying the black hole population in {\sc Dark Sage} is not enough on its own to decrease the scatter in stellar mass down to $0.2\ \mathrm{dex}$ at the low mass bins, and in fact fixing the scatter in black hole mass by overwriting {\sc Dark Sage}'s black hole masses actually tends to {\it increase} the scatter in stellar mass for a given halo mass and $\sigma_\mathrm{BH}$. We therefore turn to more extreme modifications to the AGN feedback in {\sc Dark Sage}.


\subsection{Black Holes Turn Off Cooling } \label{AGNoff_coolingoffBHM}

\begin{figure*}[t]
\centering
\includegraphics[width=1.8\columnwidth, clip]{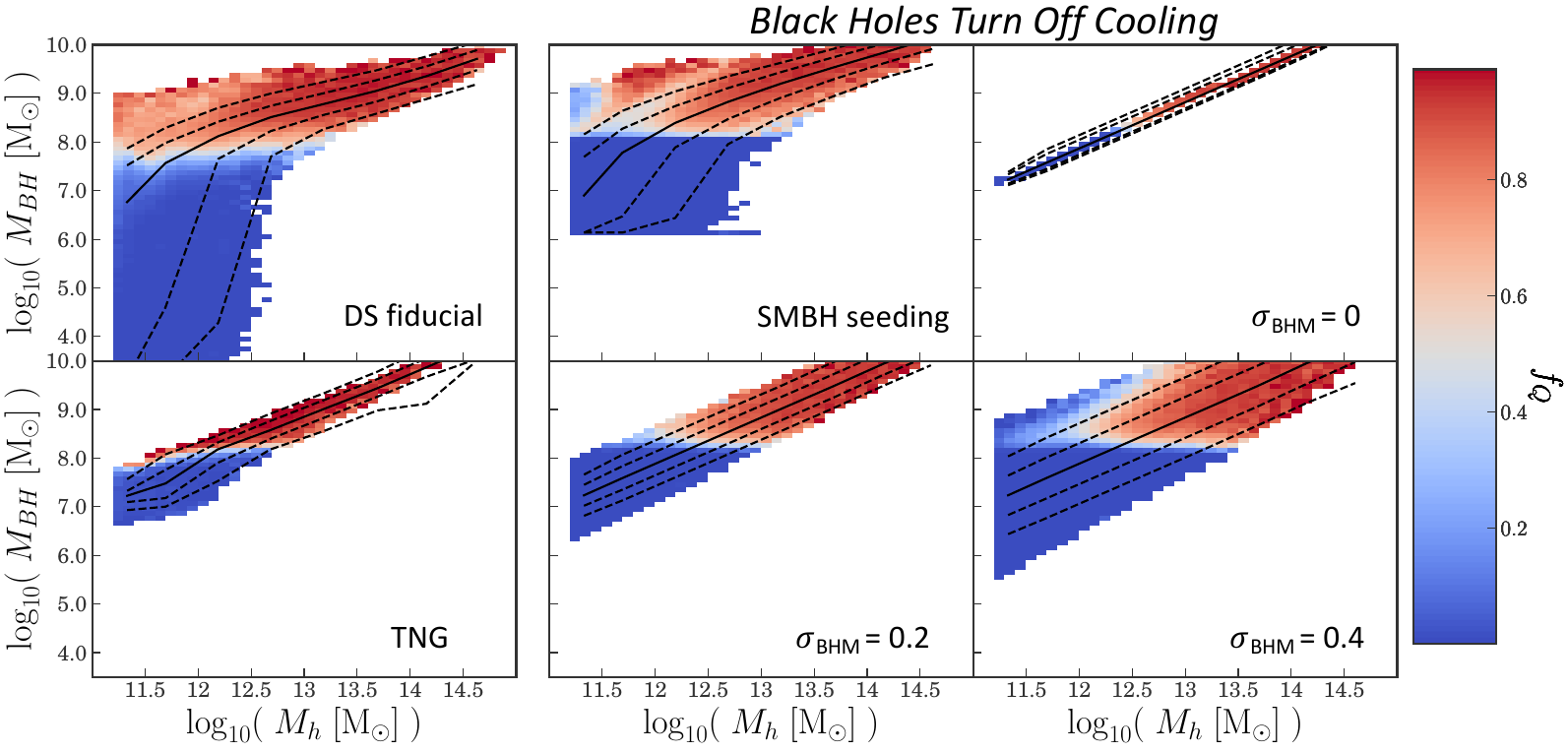}
\caption{Black hole mass as a function of halo mass colored by the quenched fraction for the same black hole models as in Figure~\ref{fig:logMvirlogBHM_classicradiomode_plot}. We see small differences in the quenching fractions in comparison to figure \ref{fig:logMvirlogBHM_classicradiomode_plot} although the black hole mass--halo mass sigma $0.2\ \mathrm{dex}$ run best resembles {\sc TNG} prediction in terms of scatter and quenched fraction transition.}
\label{fig:MBHMh_logMvirlogBHM_AGNoff_coolingoffBHM_plot}
\end{figure*}

We now introduce a new idealized AGN feedback model that enforces one of the clearest features of the {\sc TNG} black hole population, namely that galaxies with black holes more massive than $\sim 10^8 {\rm M}_\odot$ are quenched at $z=0$. In particular, rather than relying on any explicit argument about the energetics of black hole feedback, we simply turn off the radiative cooling that allows gas to accrete onto galaxies in SAMs for any galaxy with a black hole more massive than $10^8 {\rm M}_\odot$. No other form of AGN feedback operates in these runs. We specifically chose the $10^{8}\, \mathrm{M}_{\odot}$ black hole mass threshold because of the sharp quenching transition around that mass seen for example in the lower left panel of Figure \ref{fig:logMvirlogBHM_classicradiomode_plot}. In the analogous figure for these runs with black hole mass quenching, Figure \ref{fig:MBHMh_logMvirlogBHM_AGNoff_coolingoffBHM_plot}, we see some of the same features visible in Figure \ref{fig:logMvirlogBHM_classicradiomode_plot}. In particular, there are still some galaxies with high-mass black holes that continue to form stars, but they are much rarer. The black hole mass-halo mass run with $\sigma_\mathrm{BH}=0.2\ \mathrm{dex}$ looks quite similar to the {\sc TNG} black hole mass-halo mass relation in terms of scatter and transition of the quenched population when looking at the quenched fractions.

\begin{figure*}[t]
\centering
\includegraphics[width=1.8\columnwidth, clip]{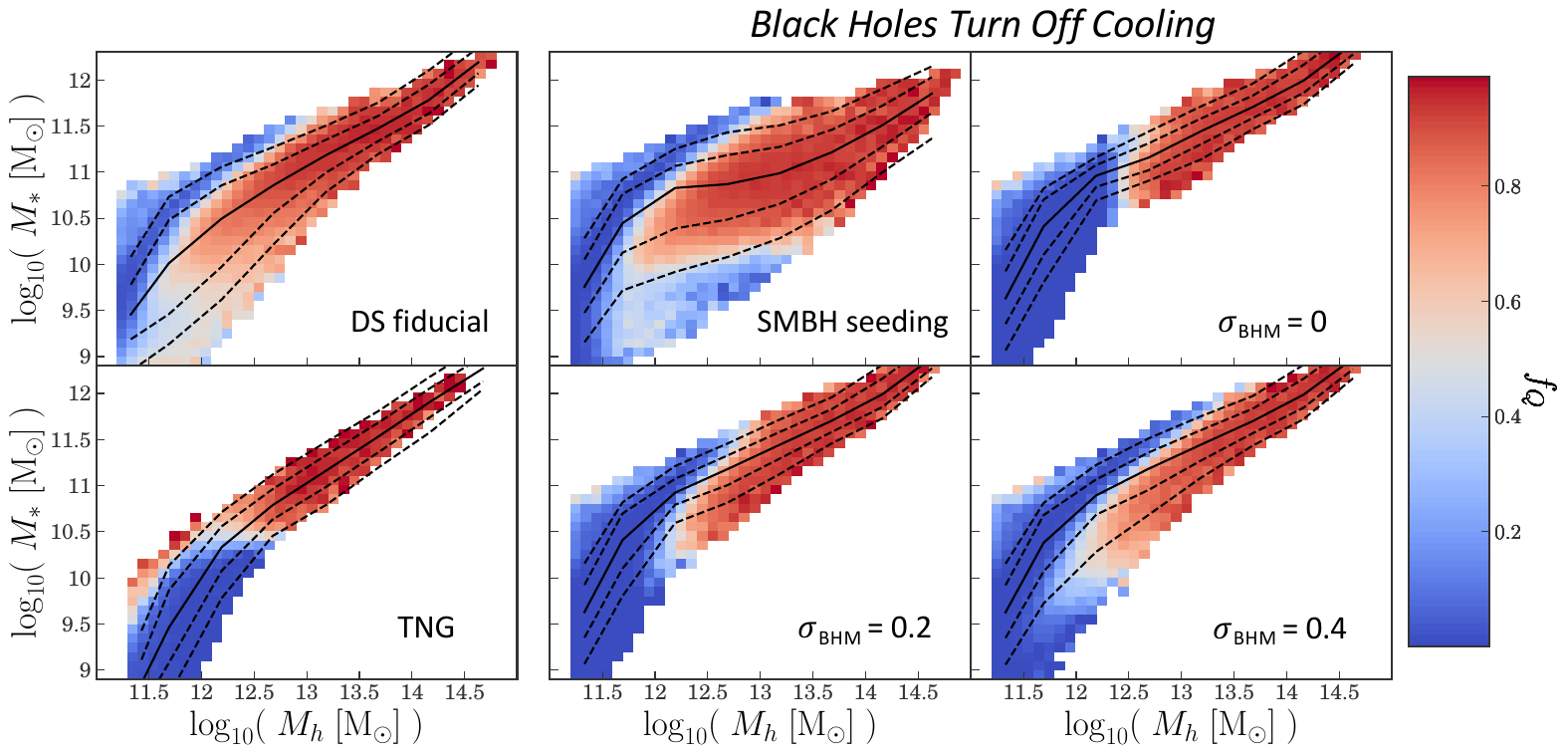}
\caption{Turning off AGN for all galaxies and turning off cooling for galaxies with black hole mass above $10^{8}\, \mathrm{M}_{\odot}$: Stellar-to-halo mass relation colored by the quenched fraction for the same black hole models as in Figure~\ref{fig:logMvirlogBHM_classicradiomode_plot}. We see a decrease in the scatter in stellar mass at fixed halo mass, more so for the conditional black hole model. There is also a clear vertical transition between active and passive galaxies, more apparent for our fixed conditional distribution models.}
\label{fig:MBHMh_SMHMR_AGNoff_coolingoffBHM_plot}
\end{figure*}

We then see how this translates to the stellar mass-to-halo mass relation in Figure \ref{fig:MBHMh_SMHMR_AGNoff_coolingoffBHM_plot}. The low-mass black hole population in massive stellar mass galaxies (as seen in Figure \ref{fig:SMHMR_classicradiomode_plot}) is still present for the SMBH seeding model and the large $\sigma_\mathrm{BH}$ runs using the fixed conditional distribution run. This is to be expected because these galaxies remain star-forming owing to their low-mass black holes in this case just as much as in the instantaneous AGN feedback case. One striking feature is the vertical dividing line in the $\sigma_\mathrm{BH}=0$ case, which appears because in this run black hole mass and halo mass have an exact 1:1 relation, so this line is where the black hole masses cross $10^8 {\rm M}_\odot$. 

\begin{figure}[t]
\centering
\includegraphics[width=\columnwidth, clip]{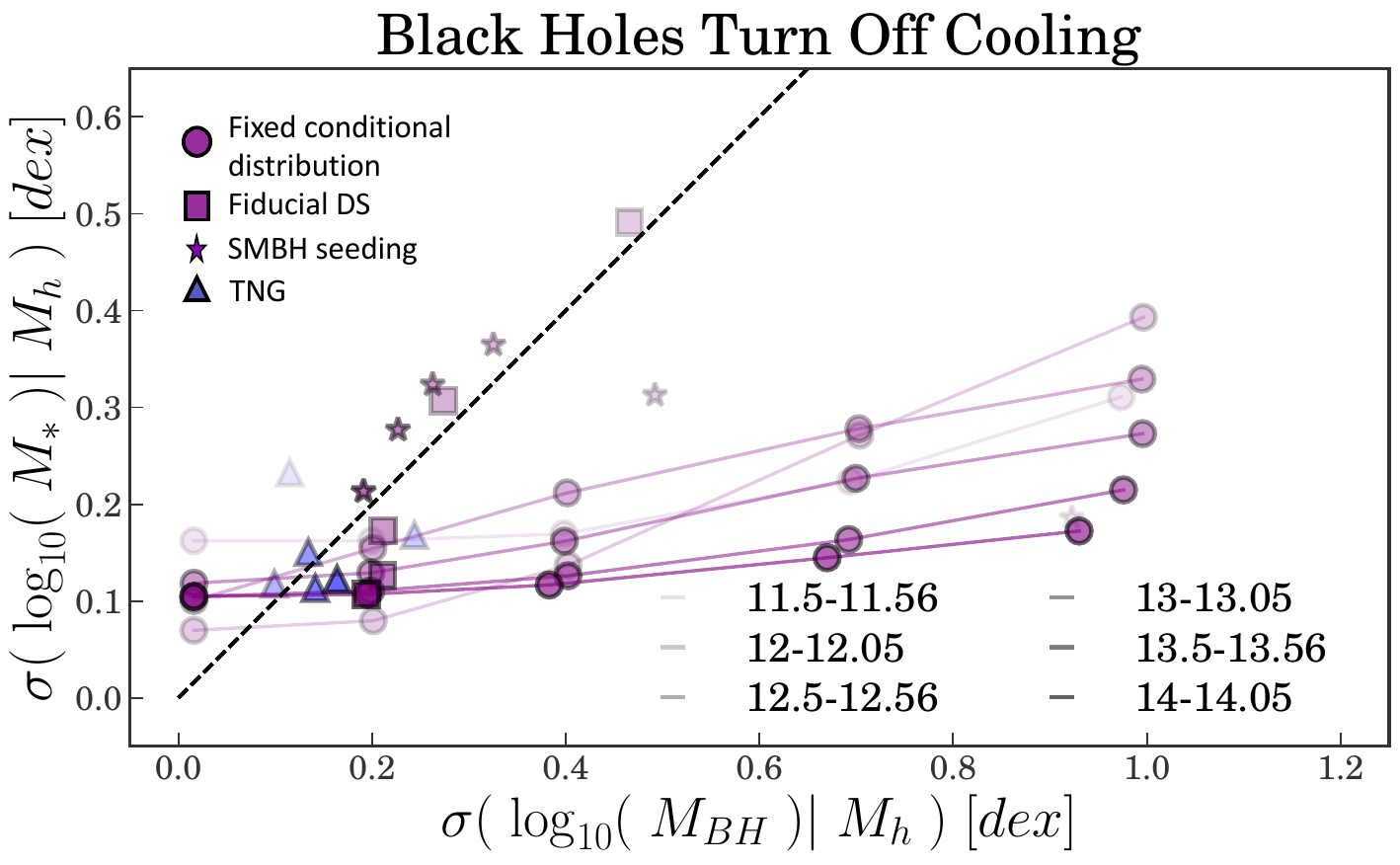}
\caption[Same as Figure \ref{fig:scatterMstarMh_scatterMBHMh_classicradiomode_plot}, except that in this model, we turn off AGN feedback and turn off cooling for galaxies with black hole masses above $10^{8}\, \mathrm{M}_{\odot}$]{Same as Figure \ref{fig:scatterMstarMh_scatterMBHMh_classicradiomode_plot}, except that in this model, we turn off AGN feedback and turn off cooling for galaxies with black hole masses above $10^{8}\, \mathrm{M}_{\odot}$. When adapting different black hole seeding, we see a significant reduction in the scatter in stellar mass at low halo mass bins.}
\label{fig:scatterMstarMh_scatterMBHMh_AGNoff_coolingoffBHM_plot}
\end{figure}

We dissect the effect of this AGN quenching model by looking at the scatter in stellar mass as a function of the scatter in black hole mass (Figure \ref{fig:scatterMstarMh_scatterMBHMh_AGNoff_coolingoffBHM_plot}). When adopting the SMBH seeding, we see a slight increase in the scatter in stellar mass at low halo mass bins. We see that, in the SMBH seeding model, some of the scatter values in stellar mass are traced by the one-to-one line, except for one value that falls around $0.4\ \mathrm{dex}$ in black hole mass scatter. For the fixed conditional distribution model, around the halo quenching mass and below, we can reduce the scatter in stellar mass by controlling for the scatter in black hole mass. The most dramatic change in the scatter in stellar mass takes place for the lowest halo mass bin, which is reduced to $0.18\ \mathrm{dex}$ of scatter in stellar mass for a black hole mass scatter of $0\ \mathrm{dex}$. For the highest halo mass bin, regardless of which black hole mass scatter value we choose, the scatter in stellar mass seems to be almost unaffected. The fixed conditional distribution can produce stellar mass scatter small enough to match {\sc TNG} predictions, at least for the lower halo mass bins.

\subsection{Black holes remove ISM} \label{AGNoff_coolingoffBHM_nocoldgasdisk}

\begin{figure*}[t]
\centering
\includegraphics[width=1.8\columnwidth, clip]{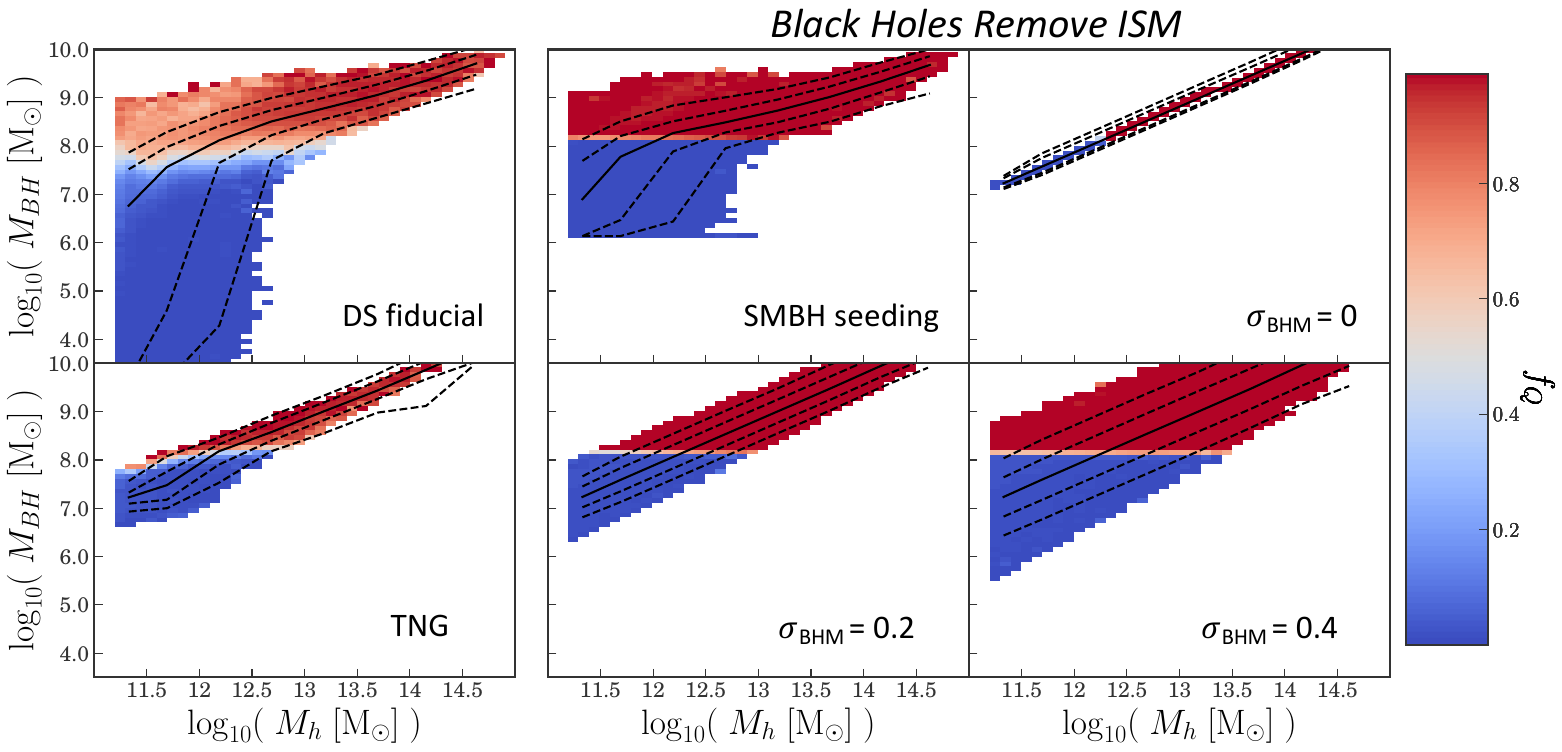}
\caption{No AGN, turning off cooling, and removing cold gas resevoir for galaxies with black hole masses above $10^{8}\, \mathrm{M}_{\odot}$: Black hole mass as a function of halo mass colored by the quenched fraction for the same black hole models as in Figure~\ref{fig:logMvirlogBHM_classicradiomode_plot}. We see the sharpest quenching transition of all AGN quenching models, where everything right above black hole masses above $10^{8}\, \mathrm{M}_{\odot}$ is quenched.}
\label{fig:MBHMh_logMvirlogBHM_AGNoff_coolingoffBHM_nocoldgasdisk_plot}
\end{figure*}

The last and most extreme idealized AGN quenching model we test is adopting the same model as in section \ref{AGNoff_coolingoffBHM} (i.e. turning off AGN feedback in {\sc Dark Sage} and cooling) while also removing all cold ISM gas for galaxies with black hole masses above $10^{8}\, \mathrm{M}_{\odot}$. Figure \ref{fig:MBHMh_logMvirlogBHM_AGNoff_coolingoffBHM_nocoldgasdisk_plot} shows the effect of this AGN quenching model for both of our black hole seeding runs. We see the sharpest quenching transition of all AGN quenching models, where everything right above black hole masses above $10^{8}\, \mathrm{M}_{\odot}$ is quenched. This is true for the SMBH seeding model as well as all the runs done using the fixed conditional distribution with different values of the dispersion.

\begin{figure*}[t]
\centering
\includegraphics[width=1.8\columnwidth, clip]{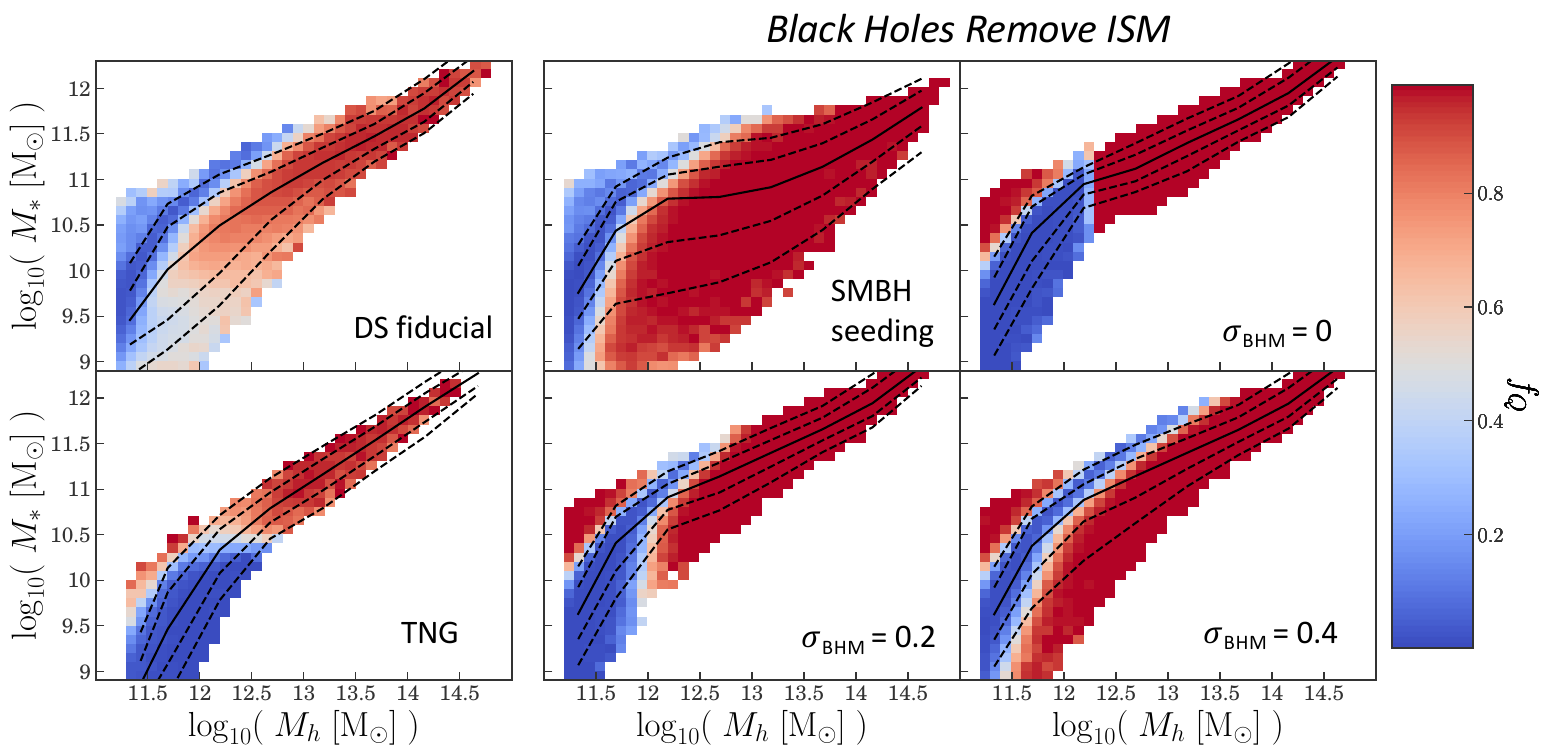}
\caption{No AGN, turning off cooling, and removing cold gas resevoir for galaxies with black hole masses above $10^{8}\, \mathrm{M}_{\odot}$: Stellar-to-halo mass relation colored by the quenched fraction for the same black hole models as in Figure~\ref{fig:logMvirlogBHM_classicradiomode_plot}. We see that the scatter in stellar mass is significantly reduced like in figure  \ref{fig:MBHMh_SMHMR_AGNoff_coolingoffBHM_plot}, but there is also less galaxies in transition from being active to passive.}
\label{fig:MBHMh_SMHMR_AGNoff_coolingoffBHM_nocoldgasdisk_plot}
\end{figure*}

We explore how this AGN quenching model affects the stellar mass-to-halo mass relation. Figure \ref{fig:MBHMh_SMHMR_AGNoff_coolingoffBHM_nocoldgasdisk_plot} shows that galaxies at all masses are significantly affected by this aggressive quenching model. We see that the scatter in stellar mass is significantly reduced like in figure \ref{fig:MBHMh_SMHMR_AGNoff_coolingoffBHM_plot}, but there are also fewer galaxies in transition from active to passive. We also see a strong strip of star-forming galaxies, especially in the large scatter values of the fixed conditional distribution, as seen in  figure  \ref{fig:MBHMh_SMHMR_AGNoff_coolingoffBHM_plot}. As mentioned before, these galaxies have low-mass black holes that have not grown significantly. 

Lastly, we take a look more closely at how this aggressive quenching affects the scatter in stellar mass when controlling for the scatter in black hole mass. As in Figure \ref{fig:scatterMstarMh_scatterMBHMh_AGNoff_coolingoffBHM_plot}, we see a significant reduction in the scatter in stellar mass at all halo mass bins, except at the highest. For the fixed conditional distribution model, around the halo quenching mass and below, the scatter in stellar mass is significantly reduced by controlling for the scatter in black hole mass. In this case, as we increase black hole mass scatter, the change in the scatter in stellar mass is comparable to Figure \ref{fig:scatterMstarMh_scatterMBHMh_AGNoff_coolingoffBHM_plot}. 

In the case of SMBH seeding, as in Figure \ref{fig:scatterMstarMh_scatterMBHMh_AGNoff_coolingoffBHM_plot}, the lowest halo mass bins have an increased stellar mass scatter of $0.3-0.4\ \mathrm{dex}$ at a fixed black hole mass scatter of $0.2\ \mathrm{dex}$. This model still predicts a lower scatter at the lowest halo mass bin than {\sc Dark Sage} fiducial. 

\begin{figure}[t]
\centering
\includegraphics[width=\columnwidth, clip]{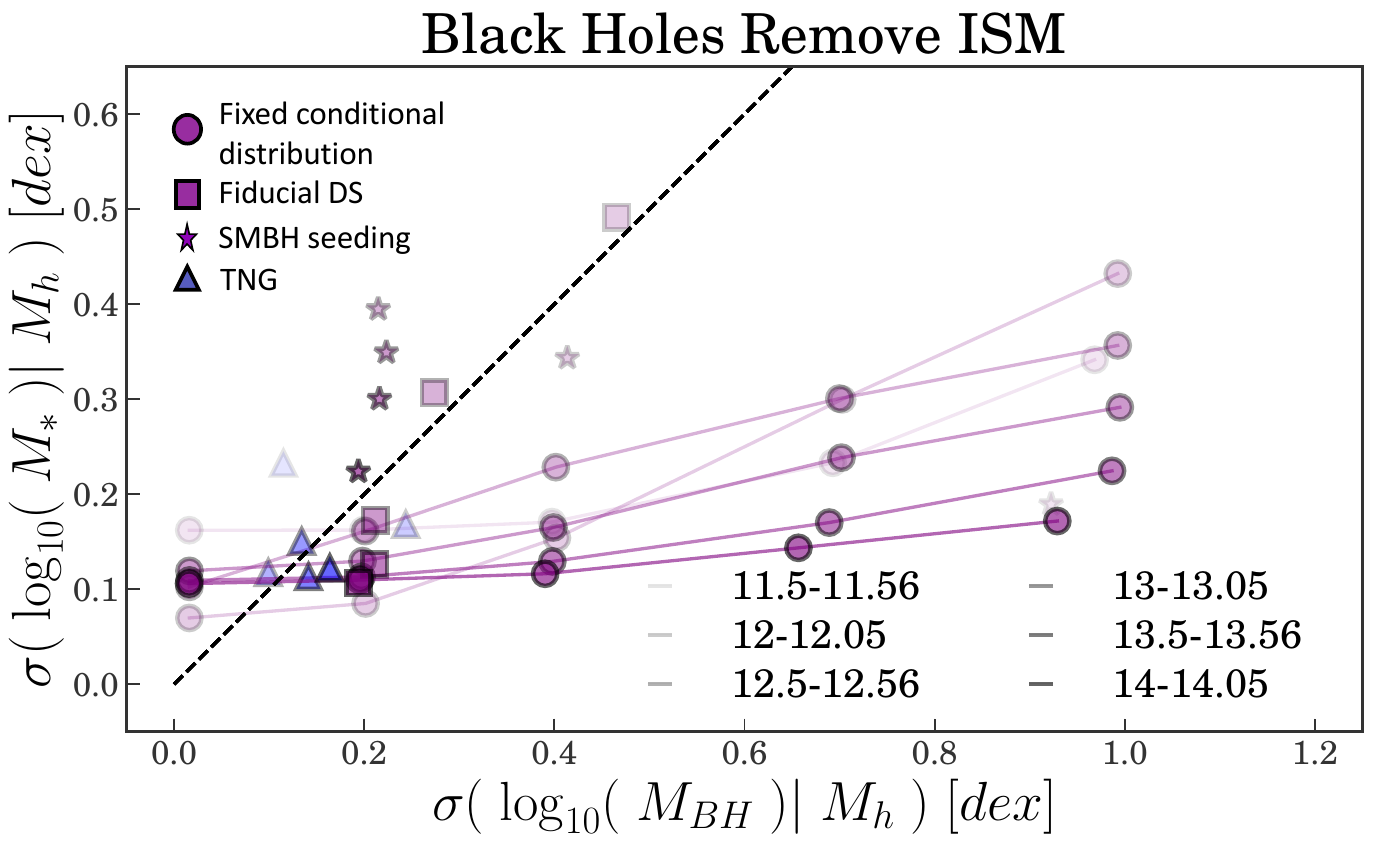}
\caption{Same as Figure \ref{fig:scatterMstarMh_scatterMBHMh_AGNoff_coolingoffBHM_plot}, except that in this model, we also remove all cold gas reservoir for galaxies with black hole masses above $10^{8}\, \mathrm{M}_{\odot}$. Although the scatter in stellar mass is significantly reduced like in figure \ref{fig:scatterMstarMh_scatterMBHMh_AGNoff_coolingoffBHM_plot}, in the SMBH seeding case, the scatter in stellar mass is slightly different for the lower halo mass bins, where the black hole mass scatter is $0.2\ \mathrm{dex}$.}
\label{fig:scatterMstarMh_scatterMBHMh_AGNoff_coolingoffBHM_nocoldgasdisk_plot}
\end{figure}

\section{Discussion}\label{discussion}

\begin{figure*}[t]
\centering
\includegraphics[width=2.0\columnwidth, clip]{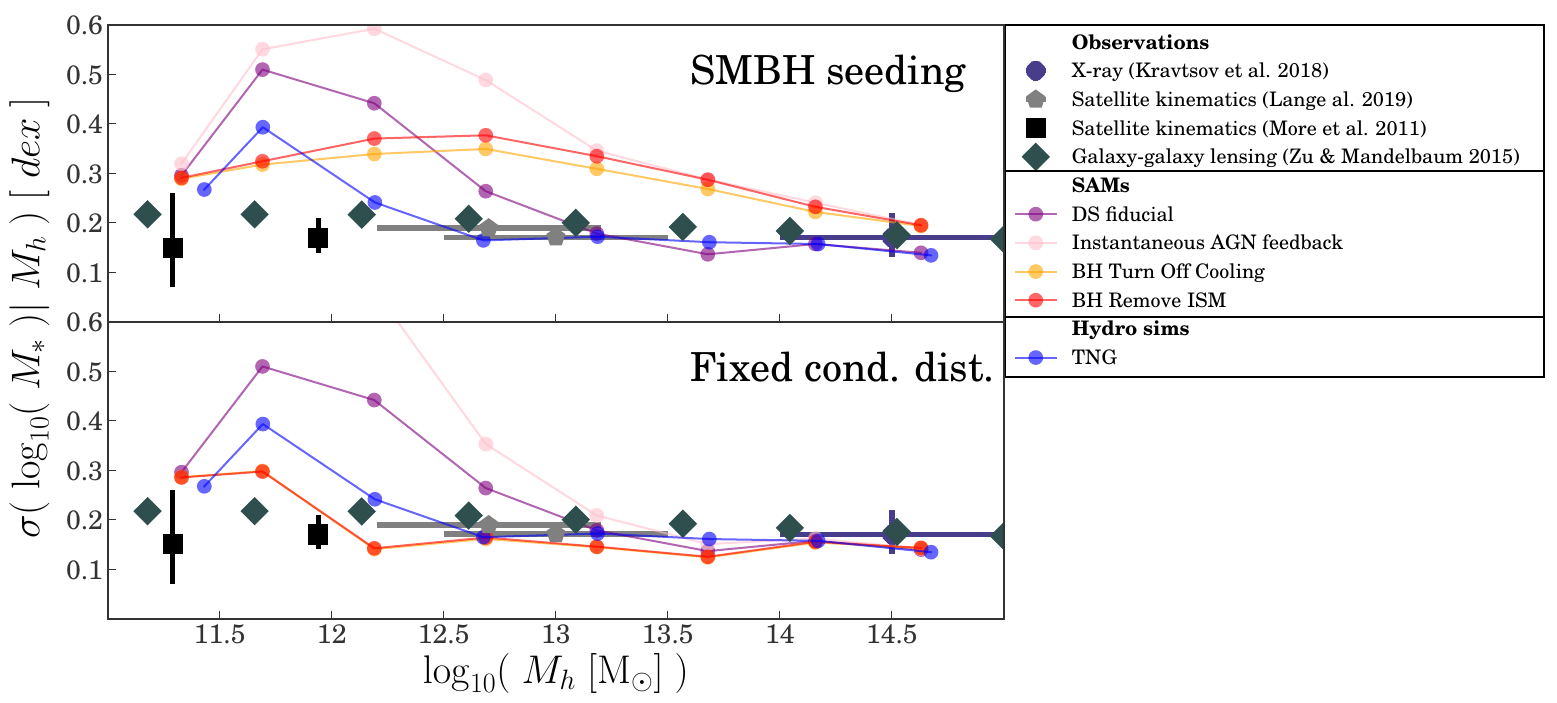}
\caption{The stellar mass scatter -- computed as half the 16th--84th interpercentile range -- at fixed halo mass for the SMBH seeding (top) and the fixed conditional distribution (bottom) BH growth models at different halo mass bins for the three different AGN feedback models: {\sc Dark Sage} instantaneous AGN feedback (pink), turning off AGN feedback while also turning of cooling for galaxies with black hole masses above $10^{8}\, \mathrm{M}_{\odot}$ (orange), and lastly, doing the former while also removing all the cold gas mass (red). We compare our experiments with observational data used in figure \ref{fig:SMHMRscatter_plot}, {\sc Dark Sage} fiducial (purple), and {\sc TNG} (blue). In the bottom panel, we only display the models with a black hole mass scatter of 0.2$\ \mathrm{dex}$.}
\label{fig:scatterTSM_BHseeding_MBHMh_DS_TNG_quenchingmodels_plot}
\end{figure*}

One of the most fundamental questions in galaxy formation relates to quantifying and understanding the origin of scatter in observable galaxy properties at a given halo mass. Observational measurements using satellite kinematics, galaxy--galaxy lensing, and X-rays yield constraints of $\sigma(\log_{10}(M_* | M_h)) 
\sim 0.2\ \mathrm{dex}$ for central galaxies with halo masses between $10^{11}$--$10^{15}\, \mathrm{M}_{\odot}$ at $z = 0$, with the blue centrals having the highest uncertainty of $_{-0.08}^{+0.11}$ (see Figure \ref{fig:SMHMRscatter_plot} or \ref{fig:scatterTSM_BHseeding_MBHMh_DS_TNG_quenchingmodels_plot}) \citep{2011MNRAS.410..210M, Zu2015MappingDR7, Kravtsov2018StellarHalos, 2019MNRAS.487.3112L}. Many empirical models assume a log-normal and constant stellar mass scatter at all halo masses \citep{2010ApJ...717..379B, 2010ApJ...710..903M}. At fixed halo mass, galaxies are assigned stellar masses randomly within a predetermined scatter, independently of other halo properties. In the case of \citet{Hearin2013TheColour, 2015arXiv150703605B}, they are consistent with the direct observational constraints and with hydrodynamical simulations {\sc TNG}, {\sc EAGLE}, and {\sc MassiveBlack II}. Interestingly, \citet{Behroozi2019Universemachine:010} disagrees with other empirical models. Instead, they find $\sigma(\log_{10}(M_* | M_h)) \sim 0.3\ \mathrm{dex}$ across all halo masses, similar to some semi-analytic model results \citep{2012MNRAS.423.1992S, 2014ApJ...795..123L, Henriques2015, Croton2016}. \citet{Behroozi2019Universemachine:010} map the star formation rate distributions to the halo assembly histories. The large scatter values could be a result of the mismatch of the correlation when galaxies are quenched versus the halo properties used at the time of quenching. SAMs systematically predict higher scatters than any other model, and the model that predicts the highest stellar mass scatter is {\sc Dark Sage}.

In the case of physical models, both semi-analytic frameworks and hydrodynamical simulations are calibrated using observed stellar mass functions, without necessarily controlling the stellar mass scatter. 
Fundamentally, the scatter in stellar mass is tied to the diverse growth of the galaxies. Dark matter halos harbor hot gas that later cools into the ISM. Over the course of the galaxy's lifetime, a fraction of the cold gas turns into stars. One mechanism that controls the mass growth in galaxies is AGN feedback. Figure \ref{fig:growthmass_heatmap_plot} shows that, after $10/\ln(10)$ Gyr, galaxies may either evolve alongside the SMHMR or away from it, contributing to an increase in the stellar mass scatter. Here, we chose three different (0.5 dex)$^2$ bins in the stellar mass-to-halo mass space. The top left bin included galaxies with high stellar masses that hosted black holes with masses $\la 10^{8}\, \mathrm{M}_{\odot}$, whereas the bottom right bin has galaxies with low stellar masses with black holes above such mass. However, just changing where black holes live is not enough to significantly reduce the scatter in stellar mass. In fact the opposite effect may happen if a high $\sigma(\log_{10}(M_{\rm BH} | M_h))$ is chosen in our fixed conditional distribution. A combination of changing black hole seeding and varying the AGN feedback in the model results a significant reduction in the scatter.

We can directly compare the effects of the experiments we performed in sections \ref{fiducial_AGNfeedback}, \ref{AGNoff_coolingoffBHM}, and \ref{AGNoff_coolingoffBHM_nocoldgasdisk} for both SMBH seeding and the fixed conditional distribution. Figure \ref{fig:scatterTSM_BHseeding_MBHMh_DS_TNG_quenchingmodels_plot} shows that adapting the SMBH seeding and: a) using {\sc Dark Sage} instantaneous AGN feedback does little to change the scatter in stellar mass, b) turning off AGN feedback while also turning off cooling for galaxies with black hole masses above $10^{8}\, \mathrm{M}_{\odot}$ reduces the scatter in stellar mass significantly for galaxies with halos below $10^{13}\, \mathrm{M}_{\odot}$. All three AGN quenching models increase the scatter in stellar mass at the highest halo mass bins. Even though the scatter in stellar mass was reduced in some cases for the lower halo mass bins, using this model, the scatter in stellar mass could not be reduced as low as observations nor as {\sc TNG} predicts.

In the fixed conditional distribution case, we find that the scatter in stellar mass can be reduced much more than in the SMBH seeding model depending on, which quenching model we choose. While the SMBH seeding model that uses {\sc Dark Sage} instantaneous AGN feedback does not significantly change the scatter in stellar mass compared to {\sc Dark Sage} fiducial, in the fixed conditional distribution case, using the instantaneous AGN feedback increases the scatter significantly, reaching stellar mass scatters as high as $0.8\ \mathrm{dex}$. However, when turning off AGN feedback while also turning off cooling for galaxies with black hole masses above $10^{8}\, \mathrm{M}_{\odot}$, the stellar mass scatter is strongly decreased, reaching values as low as $0.16\ \mathrm{dex}$ for halos around the quenching mass. The flip in the highest to lowest stellar mass scatter for the same halo mass bins when comparing the former and the latter quenching models is due to the lack of the contribution of AGN feedback to the overall galaxy population. In the case of our quenching model, we are only quenching galaxies with black hole masses above $10^{8}\ \mathrm{M}_{\odot}$, essentially not treating galaxies with black holes below that mass to have any form of quenching. We have seen that the contribution to the \textit{radio mode} feedback is so significant that by turning it off, we can get scatters lower than $0.2\ \mathrm{dex}$ below the quenching mass (see Figure \ref{fig:SMHMRscatter_plot}). Turning off AGN feedback results in galaxies around the quenching mass having lower stellar mass scatter. When comparing this quenching model with the most aggressive form of quenching, which entirely removes the cold gas reservoir from galaxies, we see that the scatter in stellar mass are equally traced between both quenching models. Yet again, an interesting finding is that none of the three AGN quenching models affects the scatter in stellar mass at the highest halo mass bins. However, with two of our quenching models, we were able to reduce the scatter in stellar mass as low as in {\sc TNG} and observations. 


Upon closer examination, none of the {\sc Dark Sage} modifications are able to reproduce the pattern of the quenched fraction in the SMHMR seen in {\sc TNG}. {\sc TNG}'s transition from mostly star-forming galaxies to mostly-quenched galaxies is sharp and occurs at a fixed stellar mass (rather than at a fixed halo mass), as can be seen by the sharp horizontal feature in plots of the SMHMR colored by $f_Q$. In our modifications to {\sc Dark Sage} and its black holes, the transitions between low-$f_Q$ and high-$f_Q$ appear closer to vertical lines, meaning that quenching is being set by the halo mass.  
Our models produce halo-mass driven quenching because black hole mass and the aggressiveness of our AGN feedback models are the primary galaxy quenching mechanisms. In {\sc TNG}, the black hole model is sometimes explicitly tied to halo mass, particularly at the stage of seeding the black holes, but as the black hole and the galaxy grow together from the same gas, the black holes' mass becomes more closely tied to the stellar mass than the halo mass. Modifying our fixed conditional distribution black hole mass model to be dependent on stellar mass may result in a stellar-mass driven quenching model, which would more closely resemble the pattern seen in {\sc TNG}.

Around the quenching mass, {\sc Dark Sage} fiducial has a larger stellar-mass scatter than any other SAMs shown in this study. The two major contributors to this are \textit{radio mode} feedback and disk instabilities (see figure \ref{fig:SMHMRscatter_plot}). From our study, we see that limiting the amount of cooling gas can reduce the scatter significantly. Most SAMs treat disk instabilities globally, accounting for all gas or stars within the galactic disk \citep{2012MNRAS.423.1992S, 2014ApJ...795..123L, Henriques2015, Croton2016}. {\sc Dark Sage} resolves disk instabilities at each of its annuli within the disk, allowing for radial dispersion of mass between adjacent annuli. The model calculates a $Q$ parameter to determine whether the gas or star disk is unstable. For the gas, this parameter \citep{Toomre1964, 1987gady.book.....B} is defined as:

\begin{equation}
\label{eq:disk_insta}
\begin{aligned}
Q_\mathrm{gas}(r) = \kappa(r)\, c_s\, /\, \pi\, G\, \Sigma_\mathrm{gas}(r),
\end{aligned}
\end{equation}
where $\kappa(r)$ is the epicyclic frequency obtained from \citet{2007asfl.book.....P} as a function of the local circular velocity and its radial derivative, $\Sigma_{\rm gas}(r)$ is the cold gas surface density, $G$ is the gravitational constant, and $c_s$ is the sound speed. {\sc Dark Sage} approximates the $c_s \simeq \sigma_{\rm gas} = 11$ km\,s$^{-1}$, which is appropriate for most nearby galaxies, but is substantially lower than observed values at higher redshift \citep[e.g.][]{2018ApJS..238...21F} and higher star formation rates \citep[e.g.][]{Green2010}. Low sound speeds result in low values of $Q$, making disks more prone to intense centrally-concentrated bursts of star formation. This catastrophic compaction of the galaxy, which is allowed in {\sc Dark Sage}, depends on another independent parameter, namely the angular momentum of newly-arriving gas. Because this is independent of the disk dynamics, i.e. it is set by accretion from the halo, we speculate that {\sc Dark Sage} galaxies experience a very wide range of outcomes, with many galaxies experiencing compaction. While such bursts are possible and may explain the existence of compact star-forming systems at $z\sim 2$ \citep[e.g.][]{Dekel2013, Barro2017}, models that allow higher velocity dispersions as the result of the disk instability itself \citep[e.g.][]{Forbes2014a} produce less-bursty evolution because some level of self-regulation of this process is allowed. The ability of the disk to self-regulate in this way may be more limited than commonly assumed \citep{Forbes2023b}, but nonetheless higher velocity dispersions driven by the instability itself are likely necessary to explain the trend between velocity dispersions and star formation rates \citep{Krumholz2018, Forbes2023a}. By turning off disk instabilities in {\sc Dark Sage}, we allow gas to remain at large galactocentric radii forming stars slowly, and therefore recover a smaller scatter in stellar mass regardless of the black hole physics. 

In addition to differences in physics between the models, there is also some uncertainty around the definition of stellar mass itself, particularly for massive elliptical galaxies. Central galaxies hosted in massive clusters contain a significant fraction of intra-cluster light in their outer regions, which sometimes is interpreted as a separate component formed from disrupted satellites and mergers. About 20-40\% of the total stellar mass in the bright cluster galaxies may come from the stellar halo \citep{2005ApJ...618..195G, 2005MNRAS.358..949Z, 2007ApJ...666..147G, 2007MNRAS.382.1940S, 2010MNRAS.403L..79M, 2013ApJ...778...14G}. While \citet{Kravtsov2018StellarHalos} includes the intra-cluster star mass in their total stellar mass, not all observational nor empirical studies explicitly state whether this component is included in their definition of stellar mass. Moreover given the large mass of the intra-cluster component, the particular division chosen between the stellar mass of the galaxy and its intra-cluster light can make a large difference in not just the scatter but the mean SMHMR \citep{Springel2018FirstClustering}. Therefore, a source of disagreement at the high mass regime may come from inconsistently comparing stellar mass scatters that may include varying portions of intra-cluster stars. Interestingly, if we did not include a modification to account for the fraction of the intra-cluster star mass included in the total stellar mass (see Appendix \ref{app:intracluster_stars_a}), {\sc Dark Sage} fiducial would have the highest scatter in stellar mass at all halo masses when comparing to all literature values, and none of the models presented here would have been able to reproduce the small scatter seen in observations and {\sc TNG} at all halo masses. The inclusion of intra-cluster stars decreased the scatter in stellar mass to the point of allowing {\sc Dark Sage} fiducial to agree with observations.

Figure \ref{fig:SMHMRscatter_plot} shows that for some models, the scatter in stellar mass reaches above $0.2\ \mathrm{dex}$ for halos below $10^{12}\, \mathrm{M}_{\odot}$. An interesting future analysis is to explore whether this increase is due to resolution effects from the simulation. 
Using zoom-in cosmological simulations, \citet{Munshi_2021} find that, in general, for all galaxies below halo masses of $10^{11}\, \mathrm{M}_{\odot}$, their stellar mass scatter remains constant around $0.3\ \mathrm{dex}$. Because these zoom-in cosmological simulations have well-resolved halos, the fact that some of these large box simulations show an exponential increase of the stellar mass scatter at low-mass halos may indicate that resolution plays an important role (see {\sc TNG300} and {\sc TNG100} in figure 8 from \citealt{Wechsler2018}).

Another interesting question revolves around the effect of our black hole mass and feedback models in the active and passive galaxy populations. \citet{2019MNRAS.484..915M} compare SDSS observations and EAGLE simulated data to show that the scatter in star formation rate at fixed stellar mass ranges between 0.3--0.5\,dex for star-forming galaxies at $z = 0$. Exploring how our model that agrees with the observed scatter in stellar mass does in terms of predicting the scatter in star formation rate for star-forming galaxies may contribute to understanding the effect of black hole mass and feedback in the star-forming population evolution. The scatter in the stellar mass--halo mass relation, the star-forming main sequence, and the mass--metallicity relation have a direct bearing on how varied the feedback can be in star-forming galaxies. For instance, \citet{2014MNRAS.443..168F} found that the scatter in the mass-loading factor had to be relatively small because the scatter of the mass--metallicity relation is also small.

\section{Conclusion}\label{conclusion}

We explore the contribution of black hole mass and AGN feedback to the stellar mass scatter within the stellar mass-to-halo mass relation using the 2018 version of the semi-analytic model {\sc Dark Sage} \citep{Stevens2018ConnectingSAGE}. We use two black hole formation models that approximate {\sc TNG300}'s black hole formation and feedback. For the first model, we seed a black hole mass of $10^{6}\, \mathrm{M}_{\odot}$ for every halo that reaches $10^{10.5} \, \mathrm{M}_{\odot}$ (section \ref{seedingBH_TNG}). For the second model, we remove any form of black hole growth that fiducial {\sc Dark Sage} implements. Instead, we force all black hole masses to follow the median black hole mass - halo mass relation from {\sc TNG300}. Additionally, we include an input scatter parameter to control the black hole mass - halo mass scatter. We do this to test whether the scatter in stellar mass could change depending on the scatter in black hole mass - halo mass. 

Our results are summarized as follows:
\begin{itemize}
    \item We find that when we turn off \textit{radio mode} feedback, the scatter in stellar mass at fixed halo mass is  significantly reduced at all halo masses.
    \item When we adopt the {\sc Dark Sage} instantaneous AGN feedback, we find that the overall scatter in stellar mass is not reduced by the change in black hole seeding for this particular feedback model. For the fixed conditional distribution, we find that by controlling for the scatter in black hole mass, the scatter in stellar mass is not reduced. On the contrary, the scatter is actually increased. 
    \item When we replace the {\sc Dark Sage} fiducial AGN feedback model with a simple model where we turn off cooling for galaxies with black hole masses above $10^{8}\, \mathrm{M}_{\odot}$, when adopting different black hole seeding, we see a significant reduction in the scatter in stellar mass at some halo mass bins, with the exception of the highest. For the fixed conditional distribution, around the halo quenching mass, we can reduce the scatter in stellar mass by reducing the scatter in black hole mass.
    \item When we turn off cooling and also remove the entire cold gas reservoir for galaxies with black hole masses above $10^{8}\, \mathrm{M}_{\odot}$, we find that the scatter in stellar mass is not significantly reduced in comparison with the previous quenching model.
    \item In all three quenching models mentioned above, we were not able to reduce the stellar mass scatter in galaxies consistently at all halo masses to the level of that seen in {\sc TNG}. 
    
\end{itemize}

\acknowledgments
\textbf{ACKNOWLEDGEMENTS}
AJP was supported by NSF grant AST-1909631 for part of the duration of this project. AJPV also thanks the Flatiron Institute, Center for Computational Astrophysics Predoctoral Fellowship. We used computational facilities from the Vanderbilt Advanced Computing Center for Research and Education (ACCRE). 
ARHS is funded through the Jim Buckee Fellowship at ICRAR-UWA.
Literature reviews for this work was made using the NASA’s Astrophysics Data System. \citet{Zu2015MappingDR7}'s line in figure \ref{fig:SMHMR_plot} was plotted using {\sc halotools} \citep{2017AJ....154..190H}. Results were produced using the IPython package \citep{Perez2007IPythonFor}, \textsc{Scipy} \citep{2020SciPy-NMeth}, \textsc{matplotlib} \citep{Hunter2007MatplotlibEnvironment}, \textsc{Astropy} \citep{Robitaille2013Astropy:Astronomy}, and \textsc{NumPy} \citep{VanDerWalt2011TheComputation}.

\clearpage

\bibliographystyle{aasjournal}
\bibliography{bibfile.bib}


\appendix
 
   
\section{A. Intra-cluster stars } 
	\label{app:intracluster_stars_a} 


There is often not a clear boundary between the stars in a galaxy and those comprising its stellar halo, particularly in massive elliptical galaxies where this intra-cluster or ``intra-halo'' material (ICS) may extend hundreds of kpc from the center of the galaxy. Indeed this material may be comparable to or even exceed the stellar mass of the galaxy \citep[e.g.][]{2005ApJ...618..195G}. Some authors avoid this difficulty by including the extended material in the definition of a galaxy's stellar mass \citep[e.g.][]{Kravtsov2018StellarHalos}, but in general most SAMs attempt to distinguish the two populations. Physically this material is likely to be dominated by disrupted satellite galaxies. In {\sc Dark Sage}, material is added to the intracluster mass (as opposed to being added to the main galaxy in a merger) when the dark matter subhalo merges with the host dark matter halo sooner than expected by an estimate of the dynamical friction timescale \citep[see][for details]{Croton2016}. While the threshold for merging vs. contributing to the ICS is adjustable in both {\sc SAGE} and {\sc Dark Sage}, it is generally not among the parameters varied in the calibration procedure.

We have found that the default procedure in {\sc Dark Sage} likely overestimates the material in the ICS relative to the main galaxy. In particular, at the high-mass end of the galaxies we consider, the ratio of mass in the ICS to the stellar mass of the galaxy was $\gg 1$, and galaxies around the Milky Way's mass could have 20-40\% of the host galaxy's stellar mass in the ICS. We therefore opted to include some portion of the ICS mass in our definition of the galaxy's mass. Galaxies with a ratio of mass in the ICS to the ``Total Stellar Mass'' (TSM) of the galaxy (which does not include any contribution from the ICS) near or exceeding some threshold have some of their mass transferred from the ICS to the TSM. Both the threshold value and the fraction of the ICS mass transferred vary linearly with $\log M_h$. The fraction of ICS mass transferred is additionally multiplied by a sigmoid function in the log of the ratio $(m_\mathrm{ics}/m_\mathrm{tsm})/\mathrm{threshold}$ scaled by 0.1, so that a higher fraction of $m_\mathrm{ics}$ is transferred for galaxies exceeding the threshold by more than a factor of a few, and a negligible quantity is transferred for galaxies below the threshold by more than a factor of a few.

    To calibrate the fractions transferred and the threshold values, we compared to the data presented in \citet{2018MNRAS.475.3348H} from the HSC survey. The authors measure the extended profiles of a large sample of elliptical galaxies, and report the ratio of the mass found within the central 10 kpc to the mass within 100 kpc. We consider this a soft upper limit for the ratio of $m_\mathrm{ics}/(m_\mathrm{ics}+m_\mathrm{TSM})$. Most definitions of the division between the galaxy and its corresponding ICS would include material beyond 10 kpc, but most light should be accounted for by the time the profile reaches 100 kpc. Neither of these statements is true for every galaxy.

\begin{figure}
      \centering 
      \includegraphics[width=\columnwidth]{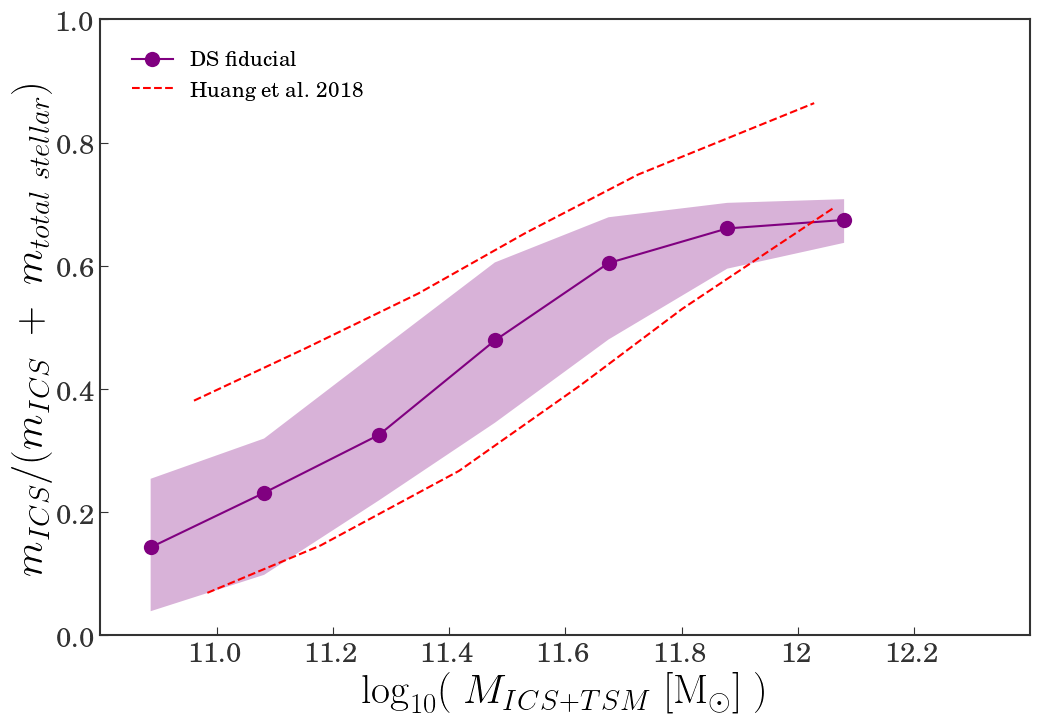}
      \caption{
          The ratio of intra-cluster stars to the total stellar mass in central galaxies at $z=0$ for the corrected {\sc Dark Sage} fiducial (purple) and the scatter. For comparison we show the range of observations of the ratio of mass of stars between 10 and 100 kpc to the stellar mass between 0 and 100 kpc for massive elliptical galaxies from \citet{2018MNRAS.475.3348H} (red dashed lines).
      }
      \label{fig:ICSfrac}
  \end{figure}

\begin{figure}
      \centering 
      \includegraphics[width=\columnwidth]{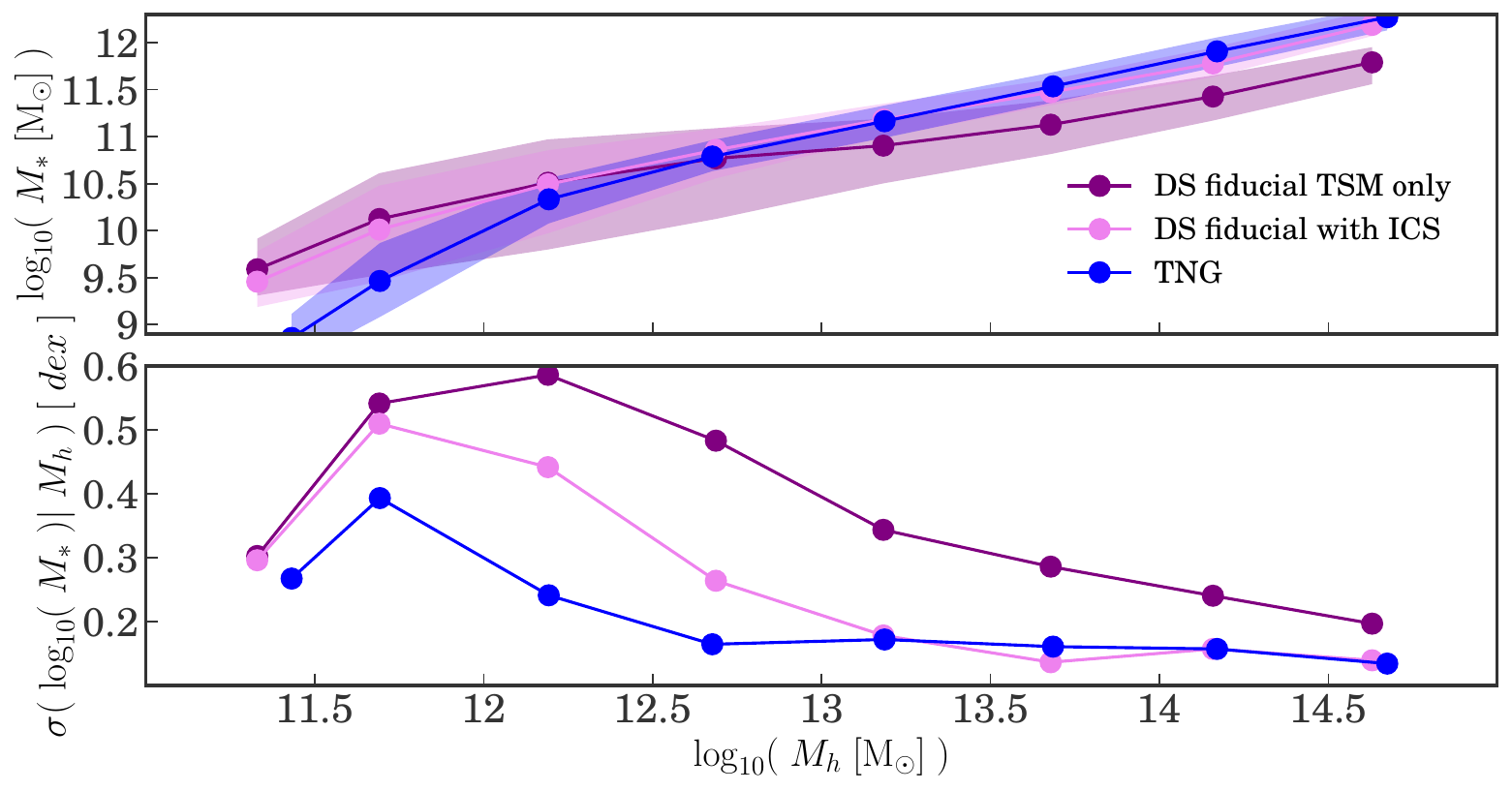}
      \caption{
          The SMHMR (top) and stellar mass scatter at fixed halo mass (bottom) for {\sc Dark Sage} fiducial, where we use the original total stellar mass of galaxies (purple), {\sc Dark Sage} fiducial, where we include a fraction between 25-100\% of the intra-cluster star mass in the stellar mass of the galaxy ($M_*$, violet), and {\sc TNG} for reference (blue). We see that not including part of the intra-cluster star mass into the total stellar mass results in galaxies at all halo masses having a higher scatter in stellar mass at fixed halo mass.
      }
      \label{fig:SMHMR_sigmastar_ICStransfer}
  \end{figure}

    We find that by varying our threshold from 0.5 to 1 from a halo mass of $10^{12} M_\odot$ to $10^{14} M_\odot$, and varying our mass transferred from 100\% at a halo mass of $10^{12} M_\odot$ down to $25\%$ at a halo mass of $10^{14} M_\odot$, we can reproduce the range of values presented by \citet{2018MNRAS.475.3348H} (see Figure \ref{fig:ICSfrac}). As a result of this adjustment, both the mean SMHMR and its scatter are affected (see Figure \ref{fig:SMHMR_sigmastar_ICStransfer}) in the direction of substantially reducing {\sc Dark Sage}'s discrepancies with TNG, particularly at very high halo masses.
    

\end{document}